\documentclass[letterpaper,11pt]{article}

\usepackage[T1]{fontenc}
\usepackage[utf8]{inputenc}

\usepackage{graphicx}
\usepackage{epstopdf}
\usepackage{multicol}
\usepackage{lmodern}

\usepackage{jheppub}
\usepackage{microtype}   
\usepackage{multirow, amsthm}
\usepackage{braket}
\usepackage{mathrsfs}
\usepackage{tikz}
\usepackage[caption=false]{subfig}
\usepackage{xspace}
\usepackage{xcolor}
\usepackage{float,color}
\usepackage[separate-uncertainty=true]{siunitx}
\usepackage[ruled,norelsize]{algorithm2e}
\usepackage[countmax]{subfloat}
\usepackage{accents}
\usepackage{physics}
\usepackage[capitalise]{cleveref}    

\usepackage{tikz}
\usetikzlibrary{arrows,shapes}
\usetikzlibrary{trees}
\usetikzlibrary{matrix,arrows} 				
\usetikzlibrary{positioning}				
\usetikzlibrary{calc,through}				
\usetikzlibrary{decorations.pathreplacing}  
\usepackage{pgffor}							

\usetikzlibrary{decorations.pathmorphing}	
\usetikzlibrary{decorations.markings}
\usetikzlibrary{snakes}
\tikzset{
    vector/.style={decorate, decoration={snake}, draw},
	provector/.style={decorate, decoration={snake,amplitude=2.5pt}, draw},
	antivector/.style={decorate, decoration={snake,amplitude=-2.5pt}, draw},
    fermion/.style={draw=black, postaction={decorate},
        decoration={markings,mark=at position .55 with {\arrow[draw=black]{>}}}},
    fermionbar/.style={draw=black, postaction={decorate},
        decoration={markings,mark=at position .55 with {\arrow[draw=black]{<}}}},
    fermionnoarrow/.style={draw=black},
    gluon/.style={decorate, draw=black,
        decoration={coil,amplitude=4pt, segment length=5pt}},
    scalar/.style={dashed,draw=black, postaction={decorate},
        decoration={markings,mark=at position .55 with {\arrow[draw=black]{>}}}},
    scalarbar/.style={dashed,draw=black, postaction={decorate},
        decoration={markings,mark=at position .55 with {\arrow[draw=black]{<}}}},
    scalarnoarrow/.style={dashed,draw=black},
    electron/.style={draw=black, postaction={decorate},
        decoration={markings,mark=at position .55 with {\arrow[draw=black]{>}}}},
	bigvector/.style={decorate, decoration={snake,amplitude=4pt}, draw},
}

\tikzstyle{block} = [draw, rectangle, 
    minimum height=3em, minimum width=6em]

\newlength{\dhatheight}

\providecommand{\href}[2]{#2}
\usepackage{comment}

\definecolor{darkred}{rgb}{0.5,0.0,0.0}
\definecolor{darkblue}{rgb}{0.0,0.0,0.9}
\definecolor{darkerblue}{rgb}{0.0,0.0,0.5}
\definecolor{darkgreen}{rgb}{0.0,0.5,0.0}
\definecolor{black}{rgb}{0.0,0.0,0.0}
\definecolor{brown}{rgb}{0.6,0.4,0.2}

\DeclareSIUnit{\nb}{\nano\barn}
\DeclareSIUnit{\pb}{\pico\barn}
\DeclareSIUnit{\fb}{\femto\barn}
\DeclareSIUnit{\year}{yr}

\def\beq{\begin{equation}}
\def\eeq{\end{equation}}
\newcommand{\bea}{\begin{eqnarray}\begin{aligned}}
\newcommand{\eea}{\end{aligned}\end{eqnarray}}

\title{\boldmath Anomaly Detection with Density Estimation
}
\author[1]{Benjamin Nachman}
\author[1,2,3]{and David Shih}
\affiliation[1]{\normalsize Physics Division, Lawrence Berkeley National Laboratory, Berkeley, CA 94720, USA}
\affiliation[2]{\normalsize NHETC, Dept. of Physics and Astronomy, Rutgers, Piscataway, NJ 08854, USA}
\affiliation[3]{\normalsize Berkeley Center for Theoretical Physics, University of California, Berkeley, CA 94720, USA}
\emailAdd{bpnachman@lbl.gov}
\emailAdd{shih@physics.rutgers.edu}

\abstract{
We leverage recent breakthroughs in neural density estimation to propose a new unsupervised anomaly detection technique (ANODE).  By estimating the conditional probability density of the data in a signal region and in sidebands, and interpolating the latter into the signal region, a fully data-driven likelihood ratio of data vs.\ background can be constructed. This likelihood ratio is broadly sensitive to overdensities in the data that could be due to localized anomalies.  In addition,  a unique potential benefit of the ANODE method is that the background can be directly estimated using the learned densities. Finally, ANODE is robust against systematic differences between signal region and sidebands, giving it broader applicability than other methods. We demonstrate the power of this new approach using the LHC Olympics 2020 R\&D Dataset. We show how ANODE can enhance the significance of a dijet bump hunt by up to a factor of 7 with a 10\% accuracy on the background prediction. While the LHC is used as the recurring example, the methods developed here have a much broader applicability to anomaly detection in physics and beyond.
}

\begin{document} 
\maketitle
\flushbottom

\section{Introduction}
\label{sec:intro}

Despite an impressive and extensive search program from ATLAS~\cite{atlasexoticstwiki,atlassusytwiki,atlashdbspublictwiki}, CMS~\cite{cmsexoticstwiki,cmssusytwiki,cmsb2gtwiki}, and LHCb~\cite{lhcbtwiki} for new particles and forces of nature, there is no convincing evidence for new phenomena at the Large Hadron Collider (LHC).   However, there remain compelling theoretical (e.g. naturalness) and experimental (e.g. dark matter) reasons for fundamental structure to be observable with current LHC sensitivity.  The vast majority of LHC searches are designed with specific signal models motivated by one of these reasons (e.g.\ gluino pair production from supersymmetry) in mind, and these searches are optimized with a heavy reliance on simulations, for both the signal and the Standard Model (SM) background. Given that it is impossible to cover every model with a specially optimized search (see e.g.~\cite{Kim:2019rhy,Craig:2016rqv} for comprehensive lists of currently uncovered models), and given that there are vast regions of unexplored LHC phase space, it is critical to consider extending the search program to include more model-agnostic methods.

A variety of model-agnostic approaches have been proposed to search for physics beyond the Standard Model (BSM) at colliders. These approaches are designed to be broadly sensitive to anomalies in data without focusing on specific models. Yet, they have varying degrees of both {\it signal-model} and {\it background-model independence}, as there is often a tradeoff between the broadness of a search and how sensitive it is to particular classes of signal scenarios.  Existing and proposed model-agnostic searches range from fully-signal-model independent but fully-background model dependent~\cite{sleuth,Abbott:2000fb,Abbott:2000gx,Abbott:2001ke,Aaron:2008aa,Aktas:2004pz,Cranmer:2005zn,Aaltonen:2007dg,Aaltonen:2007ab,Aaltonen:2008vt,CMS-PAS-EXO-14-016,CMS-PAS-EXO-10-021,Aaboud:2018ufy,ATLAS-CONF-2014-006,ATLAS-CONF-2012-107,DAgnolo:2018cun,DAgnolo:2019vbw} (because they compare data to SM simulation); to varying degrees of partial signal-model and background-model independence~\cite{Farina:2018fyg,Heimel:2018mkt,Roy:2019jae,Cerri:2018anq,Blance:2019ibf,Hajer:2018kqm,Collins:2018epr,Collins:2019jip,DeSimone:2018efk,Mullin:2019mmh,1809.02977,Dillon:2019cqt,SALAD,Aguilar-Saavedra:2017rzt}. A comprehensive overview of existing model-agnostic approaches and how they are classified in terms of signal and background model independence will be given in 
Section \ref{sec:modelindependentsearches}.

This paper introduces a new approach called \textit{ANOmaly detection with Density Estimation} (ANODE) that is complementary to existing methods and aims to be largely background and signal model agnostic.  Density estimation, especially in high dimensions, has traditionally been a difficult problem in unsupervised machine learning. The objective of density estimation is to learn the underlying probability density from which a set of independent and identically distributed examples were drawn. In the past few years, there have been a number of breakthroughs in density estimation using neural networks and the performance of high dimensional density estimation has greatly improved. The idea of ANODE is to make use of these recent breakthroughs in order to directly estimate the probability density of the data. Assuming the signal is localized somewhere, one can attempt to use sideband methods and interpolation to estimate the probability density of the background. Then, one can use this to construct a likelihood ratio generally sensitive to new physics.

As with any search for BSM, it is not enough to have a discriminant that is sensitive to signals, one must also have a valid method of background estimation, otherwise it will be impossible to claim a discovery of new physics. The method of background estimation can further introduce possible sources of signal and background model dependence, and it is important to avail oneself of data-driven background methods in any truly model-agnostic search. This paper will explore two methods of data-driven background estimation, one based on importance sampling, and the other based on directly integrating the background density estimate obtained in the ANODE procedure. 

Other neural network approaches to density estimation have been studied in high energy physics.  Such methods include Generative Adversarial Networks (GANs)~\cite{Goodfellow:2014upx,deOliveira:2017pjk,Paganini:2017hrr,Paganini:2017dwg,Butter:2019eyo,Martinez:2019jlu,Bellagente:2019uyp,Vallecorsa:2019ked,SHiP:2019gcl,Carrazza:2019cnt,Butter:2019cae,Lin:2019htn,DiSipio:2019imz,Hashemi:2019fkn,Chekalina:2018hxi,ATL-SOFT-PUB-2018-001,Zhou:2018ill,Carminati:2018khv,Vallecorsa:2018zco,Datta:2018mwd,Musella:2018rdi,Erdmann:2018kuh,Deja:2019vcv,Derkach:2019qfk,Erbin:2018csv,Erdmann:2018jxd,Urban:2018tqv}, autoencoders~\cite{Monk:2018zsb,ATL-SOFT-PUB-2018-001}, physically-inspired networks~\cite{Andreassen:2018apy,Andreassen:2019txo}, and flows~\cite{pmlr-v37-rezende15,Albergo:2019eim}.  GANs are efficient for sampling from a density and are thus promising for accelerating slow simulations, but they do not provide an explicit representation of the density itself.  For this reason, ANODE is built using normalizing flows~\cite{pmlr-v37-rezende15} and in particular the recently proposed masked autoregressive flow (MAF)~\cite{NIPS2017_6828}.   These methods estimate densities by using a succession of neural networks to gradually map the original data to a transformed dataset that follows a simple distribution (e.g.\ normal or uniform). 

The ANODE method is demonstrated using a simulated large-radius dijet search based on the LHC Olympics 2020 R\&D dataset \cite{gregor_kasieczka_2019_2629073}.  In particular, properties of hadronic jets are used as discriminating features to enhance a bump hunt in the invariant mass of pairs of jets.  ANODE learns a parameterized density of the features using a sideband and this is combined with a density estimation of the same features in the signal region.  The resulting likelihood ratio is able to enhance  the sensitivity of a traditional bump hunt from $S/\sqrt{B}\sim 1$ to $S/\sqrt{B}\gg 5$.
There is currently no dedicated search for generic dijet signatures where each of the jets can also originate from a BSM resonance~\cite{Kim:2019rhy,Aguilar-Saavedra:2017zuc,Aguilar-Saavedra:2019adu,Agashe:2018leo,Agashe:2017wss}.  Therefore, this particular application could be directly useful for extending the LHC physics search program.  Many other applications to resonant new physics searches involving jets and other final states are also possible.

In order to benchmark the performance of ANODE, it is compared with the CWoLa hunting method~\cite{Collins:2018epr,Collins:2019jip}.  The CWoLa approach is also a neural network-based resonance search, but does not involve density estimation.  Instead, CWoLa hunting uses neural networks to identify differences between signal regions and neighboring sideband regions.  By turning the problem into a supervised learning task~\cite{Metodiev:2017vrx}, CWoLa is able to effectively find rare resonant signals.  However, CWoLa hunting has certain requirements on the independence of the discriminating features and the resonant feature.  ANODE does not have this requirement and the potential for exploiting correlated features is studied by introducing correlations. 

This paper is organized as follows.  Section~\ref{sec:modelindependentsearches} reviews the landscape of model independent searches at the LHC to provide context for the ANODE method.  Section~\ref{sec:methods} introduces the details of the ANODE approach and provides a brief introduction to normalizing flows.   The reminder of the paper illustrates ANODE through an example based on a dijet search using jet substructure.  Details of the simulated samples are provided in Sec.~\ref{sec:sim} and the results for the signal sensitivity and background specificity are presented in Sec.~\ref{sec:sens} and~\ref{sec:back}, respectively.   A study of correlations between the discriminating features and the resonant feature is in Sec.~\ref{sec:correlations}.  The paper ends with conclusions and outlook in Sec.~\ref{sec:conclusions}.

\section{An Overview of Model (In)dependent Searches}
\label{sec:modelindependentsearches}

\begin{figure}[h!]
\centering
\includegraphics[width=0.95\textwidth]{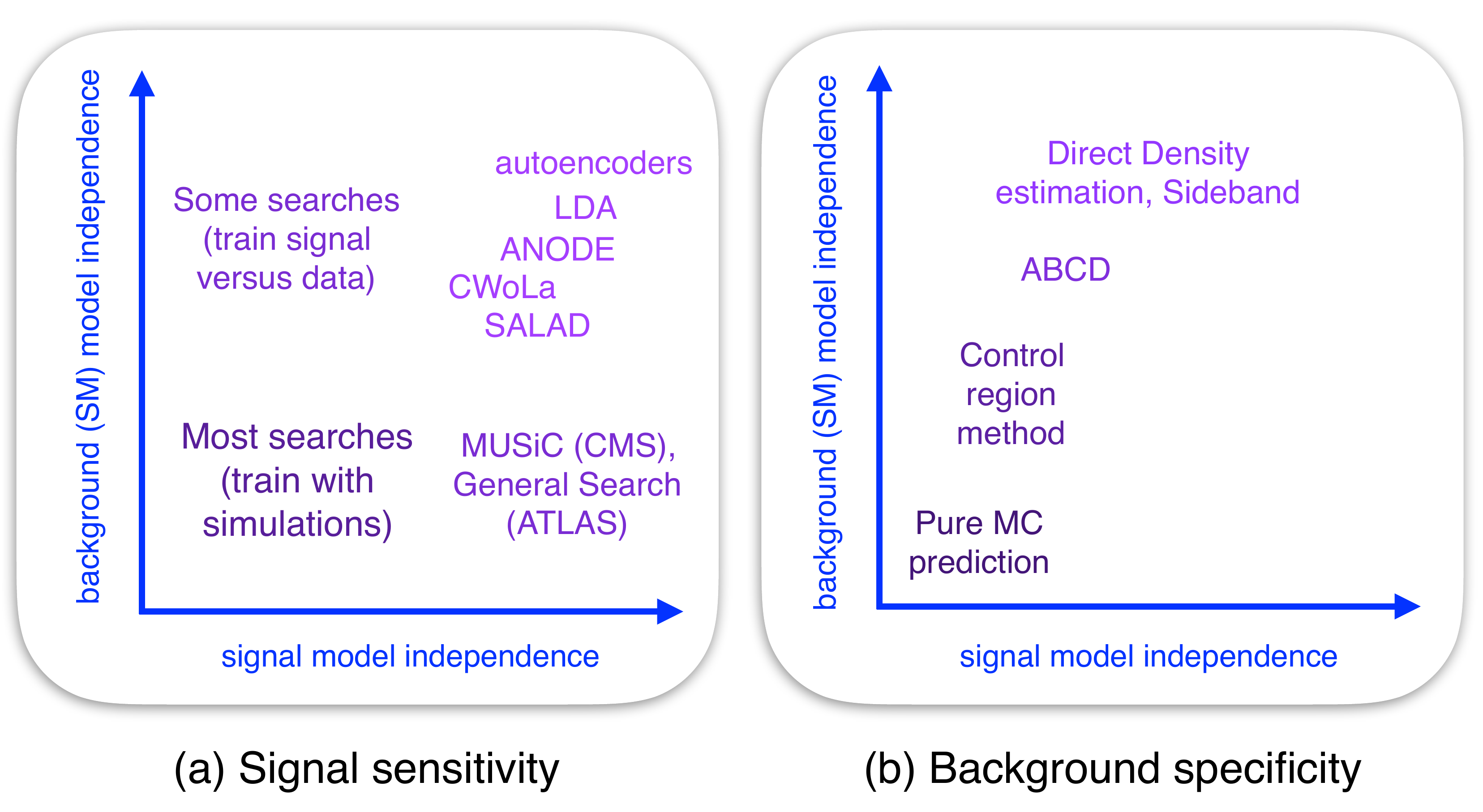}
\caption{A graphical representation of searches for new particles in terms of the background and signal model dependence for achieving signal sensitivity (a) and background specificity (b).  The Model Unspecific Search for New Physics (MUSiC)~\cite{CMS-PAS-EXO-14-016,CMS-PAS-EXO-10-021} and General Search~\cite{Aaboud:2018ufy,ATLAS-CONF-2014-006,ATLAS-CONF-2012-107} strategies are from CMS and ATLAS, respectively.   LDA stands for \textit{Latent Dirichlet Allocation}~\cite{10.5555/944919.944937,Dillon:2019cqt}, \textit{ANOmaly detection with Density Estimation} (ANODE) is the method presented in this paper, CWoLa stands for \textit{Classification Without Labels}~\cite{Metodiev:2017vrx,Collins:2018epr,Collins:2019jip} and SALAD stands for \textit{Simulation Assisted Likelihood-free Anomaly Detection}~\cite{SALAD}.  Direct density estimation is a form of side-banding where the multidimensional feature space density is learned conditional on the resonant feature (see Sec.~\ref{sec:methodbackground}).
}
\label{fig:schematic}
\end{figure}

A viable search for new physics generally must have two essential components: it must be sensitive to new phenomena and it must also be able to estimate the background under the null hypothesis (Standard Model only). The categorization of a search's degree of model (in)dependence requires consideration of both of these components. Figure~\ref{fig:schematic} illustrates how to characterize model independence for both BSM sensitivity and SM background specificity.  We will now consider each in turn.

\subsection{BSM sensitivity}

For BSM sensitivity, the various types of searches are categorized as follows:
\begin{itemize}
\item Almost all searches at the LHC are optimized (with or without machine learning) using simulations of both the SM and particular signal models.  This is represented as the lower-left corner of Fig.~\ref{fig:schematic}(a).  

\item A handful of searches use signal simulation and unlabeled data to optimize the event selection.  These are \textit{background model agnostic} and are depicted in the upper-left corner of Fig.~\ref{fig:schematic}(a).  For example, this was used in the $\gamma\gamma$ channel of the recent $t\bar{t}h$ observation, using events with inverted selection criteria to define the background data sample for optimization~\cite{Aaboud:2018urx,Sirunyan:2018ouh}.  

\item A series of \textit{signal model agnostic}, but \textit{background model-dependent} searches have been performed by D0~\cite{sleuth,Abbott:2000fb,Abbott:2000gx,Abbott:2001ke}, H1~\cite{Aaron:2008aa,Aktas:2004pz}, ALEPH~\cite{Cranmer:2005zn}, CDF~\cite{Aaltonen:2007dg,Aaltonen:2007ab,Aaltonen:2008vt}, CMS~\cite{CMS-PAS-EXO-14-016,CMS-PAS-EXO-10-021}, and ATLAS~\cite{Aaboud:2018ufy,ATLAS-CONF-2014-006,ATLAS-CONF-2012-107}.  All of these searches share essentially the same approach: they compared histograms of data to histograms of SM simulations and looked for discrepancies.  Such searches are represented in the lower-right part of Fig.~\ref{fig:schematic}(a).  Recently, there have been proposals to extend these searches with deep learning~\cite{DAgnolo:2018cun,DAgnolo:2019vbw}.

\item More recently, a variety of approaches have been proposed, often relying on sophisticated deep learning techniques, that attempt to be both signal and background model agnostic, to varying degrees. These include approaches based on autoencoders~\cite{Farina:2018fyg,Heimel:2018mkt,Roy:2019jae,Cerri:2018anq,Blance:2019ibf,Hajer:2018kqm}, weak supervision~\cite{Collins:2018epr,Collins:2019jip}, nearest neighbor algorithms~\cite{DeSimone:2018efk,Mullin:2019mmh,1809.02977}, probabilistic modeling~\cite{Dillon:2019cqt}, reweighted simulation~\cite{SALAD}, and others~\cite{Aguilar-Saavedra:2017rzt}.  These are indicated in the upper-right corner of Fig.~\ref{fig:schematic}(a).
\end{itemize}

In the upper-right corner of Fig.~\ref{fig:schematic}(a), we have also attempted to illustrate in finer detail the differences between some recent model-agnostic approaches. For example, the autoencoder is in the farthest corner since it assumes almost nothing about the signal or the background but can be run directly on the data, as long as the signal is sufficiently rare~\cite{Farina:2018fyg,Heimel:2018mkt}. The tradeoff is that there is no optimality guarantee for the autoencoder -- any signals that it does find will be found in a rather uncontrolled manner. Meanwhile, CWoLa hunting~\cite{Collins:2018epr,Collins:2019jip}  is somewhat more signal and background model-dependent than autoencoders, since this approach assumes that the signal is localized in a particular feature, and that there is an uncorrelated set of additional features on which one can train a classifier to distinguish signal region and sideband. In return, one obtains a guarantee of asymptotic optimality -- the classifier approaches the likelihood ratio~\cite{neyman1933ix} in the limit of infinite statistics\footnote{See Appendix~\ref{sec:optimal} for more details about `optimality'. }.

The ANODE method introduced in this paper complements the other recently proposed techniques and is asymptotically optimal.  To do this, ANODE estimates the density of the background-only scenario using sidebands and compares that with the density estimated in a signal-sensitive region (details are in Sec.~\ref{sec:methods}).  Like the CWoLa hunting method, the new approach is broadly sensitive to resonant new physics and thus it is placed in the upper right part of Fig.~\ref{fig:schematic}(a). 
The reason that ANODE is further right and above of CWoLa hunting is that it is less sensitive to correlations, a feature that is discussed more below.

\subsection{Background estimation}

A variety of methods are commonly used for background estimation and are highlighted in Fig.~\ref{fig:schematic}(b).   Generally, background estimation is less dependent on the signal model than achieving signal sensitivity and therefore the $x$-axis range of Fig.~\ref{fig:schematic}(b) is more compressed than Fig.~\ref{fig:schematic}(a). 
\begin{itemize} 
\item In some cases, the simulation is used to directly estimate the background.  This is often the case for well-understood backgrounds such as electroweak phenomena or very rare processes that are difficult to constrain with data.  

\item Most searches use data in some way to constrain the background prediction.  One common approach is the {\it control region method}, where a search is complemented by an auxiliary measurement to constrain the simulation.  Knowledge of the signal is used to ensure that the auxiliary measurement is not biased by the presence of signal.  

\item The two most common methods for background estimates that do not directly use simulation are the {\it ABCD method} and the {\it sideband method} (bump hunt).  The ABCD method operates by identifying two independent features, each which is sensitive to the presence of signal.  Four regions, labeled A,B,C, and D are constructed by (anti)requiring a threshold on the two features.  The background rate in the most signal sensitive region is estimated from the other three regions.  Background simulations are required to verify independence of the two features.  

\item Finally, the sideband fit only requires that the background be smooth in the region of a potential signal so that a parametric (or not~\cite{Frate:2017mai}) function can be fit to sidebands and interpolated.  However, this method only works for resonant new physics.  
\end{itemize}

While strategies from Fig.~\ref{fig:schematic}(a) can often be matched with any approach in Fig.~\ref{fig:schematic}(b), there is often one combination that is used in practice.  Table~\ref{tab:searches} provides examples of various searches and the background estimation technique that typically is associated with that search.  Searches with a complex background may use multiple background estimation procedures. 

ANODE can be combined with any background estimation technique, but it can also be used directly since the background density is already estimated to construct a signal-sensitive classifier.  Even though directly providing an accurate background estimation puts stringent requirements on the accuracy of the density estimation, it also reduces the need for a full decorrelation between classification features and the resonant feature.  A variety of decorrelation techniques exist~\cite{Louppe:2016ylz,Dolen:2016kst,Moult:2017okx,Stevens:2013dya,Shimmin:2017mfk,Bradshaw:2019ipy,ATL-PHYS-PUB-2018-014,DiscoFever,Xia:2018kgd,Englert:2018cfo,Wunsch:2019qbo}, but ultimately decorrelating removes information available for classification.  

\begin{table}[h!]
\centering
  \begin{tabular}{| c | c |c |}
    \hline
       Search & Typical Background Strategy & Recent Examples \\ \hline
    MUSiC \& the General Search  & Pure MC Prediction&~\cite{Aaboud:2018ufy,CMS-PAS-EXO-14-016} \\
    Pure electroweak processes  & Pure MC Prediction & ~\cite{Aaboud:2017rel} \\
        SUSY with top quarks \& $W$ bosons  & Control Region Method&~\cite{Aaboud:2017aeu,CMS-PAS-SUS-19-009} \\ 
        All-hadronic searches & ABCD Method &~\cite{Aaboud:2017hdf,Sirunyan:2018rlj}\\ 
        Long-lived particle searches & ABCD Method &~\cite{Aaboud:2018aqj,Sirunyan:2018vlw} \\ 
        BSM resonance searches & Sideband Method&~\cite{Aad:2019hjw,Sirunyan:2019vgj} \\ 
        CWoLa hunting & Sideband Method &~\cite{Collins:2018epr,Collins:2019jip}\\ 
        ANODE & Sideband or Direct Density & This paper\\ 
    \hline
  \end{tabular}
  \caption{A table with the common pairings of search strategy for signal sensitivity (left column), the background estimation method (middle column), and an example search (right column).}
  \label{tab:searches}
\end{table}

\section{The ANODE Method}
\label{sec:methods}

This section will describe the ANODE proposal for an unsupervised method to search for resonant new physics using density estimation. 

Let $m$ be a feature in which a signal (if it exists) is known to be localized around some $m_0$.  The value of $m_0$ will be scanned for broad sensitivity and the following procedure will be repeated for each window in $m$.  It is often the case that the width of the signal in $m$ is fixed by detector properties and is signal model independent.  A region $m_0\pm\delta$ is called the signal region (SR) and $m\not\in[m_0-\delta,m_0+\delta]$ is defined as the sideband region (SB). A traditional, unsupervised, model-agnostic search is to perform a bump hunt in $m$, using the SB to interpolate into the SR in order to estimate the background.

Let $x\in\mathbb{R}^d$ be some additional discriminating features in which the signal density is different than the background density.  If we could find the region(s) where the signal differs from the background and then cut on $x$ to select these regions, we could improve the sensitivity of the original bump hunt in $m$. The goal of ANODE is to accomplish this in an unsupervised and model-agnostic way, via density estimation in the feature space $x$. 

More specifically, ANODE attempts to learn two densities: $p_\text{data}(x|m)$ and $p_\text{background}(x|m)$ for $m\in {\rm SR}$.  Then, classification is performed with the likelihood ratio
\beq
\label{eq:Rx}
R(x|m)=\frac{p_\text{data}(x|m)}{p_\text{background}(x|m)}.
\eeq
In the ideal case that $p_\text{data}(x|m)=\alpha\, p_\text{background}(x|m)+(1-\alpha)\,p_\text{signal}(x|m)$ for $0\leq\alpha\leq 1$ and $m\in\text{SR}$, Eq.~\ref{eq:Rx} is the optimal test statistic for identifying the presence of signal.  In the absence of signal, $R(x|m)=1$, so as long as $p_\text{signal}(x|m)\neq p_\text{background}(x|m)$, $R_\text{data}(x|m)$ has a non-zero density away from 1 in a region with no predicted background.

In practice, both $p_\text{data}(x|m)$ and $p_\text{background}(x|m)$ are approximations and so $R(x|m)$ is not unity in the absence of signal.  The densities $p(x|m)$ are estimated using conditional neural density estimation as described in Sec.~\ref{sec:densityestimation}.  The function $p_\text{data}(x|m)$ is estimated in the signal region and the function $p_\text{background}(x|m)$ is estimated using the sideband region and then interpolated into the signal region.  The interpolation is done automatically by the neural conditional density estimator.  Effective density estimation will result in $R(x|m)$ in the SR that is localized near unity and then one can enhance the presence of signal by applying a threshold $R(x|m)>R_\text{cut}$, for $R_\text{cut}>1$.  The interpolated $p_\text{background}(x|m)$ can then also be used to estimate the background, as described in Sec.~\ref{sec:methodbackground}.

\subsection{Neural Density Estimation}
\label{sec:densityestimation}

The ANODE procedure as described in the previous subsection is completely general with regards to the method of density estimation. In this work we will demonstrate a proof-of-concept using normalizing flow models for density estimation. Since normalizing flows were proposed in Ref.~\cite{pmlr-v37-rezende15}, they have generated much activity and excitement in the machine learning community, achieving state-of-the-art performance on a variety of benchmark density estimation tasks. 

The core idea behind a normalizing flow is to apply a change of variables from a random variable with a simple density (e.g. Gaussian or uniform) to one with a complex density that matches some training dataset.  The transformation from one density describing random variable $X$ to another density describing random variable $Y$ follows the usual change of variables formula using the Jacobian:
\beq
\label{eq:changeofvariables}
p_Y(y)=p_X(x)\,\bigg|\text{det}\left(\frac{\partial f}{\partial x}\right)\bigg|^{-1},
\eeq
where $x$ and $y$ are realizations of $X$ and $Y$, respectively, $X$ and $Y$ have the same dimension, and $Y=f(X)$ is an invertible function.   The process in Eq.~\ref{eq:changeofvariables} can be repeated to build a normalizing flow:
\beq
\label{eq:changeofvariables2}
p_Y(y)=p_X(x)\,\prod_{i=1}^N\,\bigg|\text{det}\left(\frac{\partial f_i}{\partial y_{i-1}}\right)\bigg|^{-1},
\eeq
where $Y_i=f_i(Y_{i-1})$, $Y_0=X$, and $Y=f_N(Y_{N-1})$.   The first neural density estimation with normalizing flows had the following form for $x\in\mathbb{R}^n$:
\beq
\label{eq:originalflow}
f(x)=x+\bar{x}\,\sigma(w\cdot x + b),
\eeq
where $\sigma$ is an element-wise non-linearity and $\bar{x}\in\mathbb{R}^n, w\in\mathbb{R}^n, b\in\mathbb{R}$ are trainable parameters.  The benefit of Eq.~\ref{eq:originalflow} is that the Jacobian evaluation is simple from the chain rule.  Since the first development of normalizing flows, there has been significant development in extending their expressivity.   One innovation is to combine flows with autoregressive density estimation~\cite{NIPS2016_6581}.  An autoregressive flow~\cite{JMLR:v17:16-272}
modifies the change of variables so that for $Y_i=f(X_i)$, $Y_{i,\alpha}=f_{i,\alpha}(X_{i,1},...,X_{i,\alpha})$, where the indices $\alpha$ denote the dimension of $X_i$ and $Y_i$ for $\alpha=1,...,n$.  Any $f$ that satisfies this condition is amenable to neural density estimation because the Jacobian determinant evaluation is simple.  In particular, the Jacobian is upper triangular and therefore the determinant is the product of the diagonal elements: $\prod_{\alpha=1}^n \partial f_{i,\alpha}/\partial x_\alpha$.  ANODE is built on a masked autoregressive flow (MAF)~\cite{NIPS2017_6828}.  For a MAF, 
\beq
\label{eq:maf}
Y_{i,\alpha}=\mu_{i,\alpha}(Y_{i,1},...,Y_{i,\alpha-1})+\sigma_{i,\alpha}(Y_{i,1},...,Y_{i,\alpha-1})X_{i,\alpha},
\eeq
where $\sigma_{i,\alpha}>0$ and $\mu_{i,\alpha}$ are arbitrary functions and $Y_{i,1}=\mu_{i,1}+\sigma_{i,1} X_{i,1}$ for arbitrary numbers $\sigma_{i,1}>0$, $\mu_{i,1}$.  As in Eq.~\ref{eq:changeofvariables2}, this procedure is repeated multiple times to build a deep autoregressive flow.  The Masking in MAF comes from its use of MADE~\cite{pmlr-v37-germain15} to evaluate $\mu_{i,\alpha}$ and $\sigma_{i,\alpha}$ for all $\alpha$ in one forward pass. 
This approach eliminates the need for the recursion in Eq.~\ref{eq:maf}.  MAF is nearly the same as inverse autoregressive flows (IAF)~\cite{NIPS2016_6581}, which also use Gaussian autoregressions and are built on MADE.  The main difference is that MAF is very efficient for density estimation and slow for sampling while IAF is slow for density estimation and fast for sampling.  As ANODE only needs to estimate the density without producing new samples, MAF is selected as the method of choice.

The estimation of $p_\text{background}(x|m)$ for ANODE requires that the MAF provides a conditional density.  This can be accomplished by adding $m$ as an input to all functions $\mu_i$ and $\sigma_i$. 

\subsection{Estimating the Background}
\label{sec:methodbackground}

An anomaly detection technique is only useful for finding new particles if the Standard Model background can be estimated.  As mentioned earlier, one benefit of the direct density estimation in ANODE is that the background can be directly estimated with $p_\text{background}(x|m)$.  This results in multiple possibilities for background estimation that are considered in this work:

\begin{itemize}
\item {\it Direct density estimation}.  These methods use the interpolated $p_\text{background}(x|m)$ to directly compute the efficiency $\epsilon_{bg}(R_c|m)$ of the background after a threshold requirement on $R(x|m)$.  
\begin{itemize}
	\item Density sampling.  One could directly sample events from $p_\text{background}(x|m)$ using the stacked change of variables specified by Eq.~\ref{eq:maf}.   As mentioned in Sec.~\ref{sec:densityestimation}, this is less efficient for MAF compared with IAF.  This sampling is not pursued in this paper.
	\item Density integration.  Another approach is to directly integrate $p_\text{background}(x|m)$ for events with $R(x|m)>R_c$:
\bea\label{eq:bgesti}
\epsilon_{bg}(R_c|m) &= \int dx\,p_\text{background}(x|m)\Theta(R(x|m)-R_c). \\
\eea
	\item Importance sampling.  Analytically integrating a function in high dimensions is impractical, so one can estimate the integral with importance sampling.  An effective method to implement this sampling is make the following observation:
\bea\label{eq:bgestii}
\epsilon_{bg}(R_c|m) &=\int dx\,p_\text{background}(x|m)\Theta(R(x|m)-R_c) \\
&= \int dx\,p_\text{full}(x|m) {1\over R(x|m)} \Theta(R(x|m)-R_c)\\
&= \int_{R_c}^\infty dR\, p_\text{full}(R|m) \,{1\over R}.
\eea
The last line in Eq.~\ref{eq:bgestii} can be estimated by computing the fraction of events in the SR (representing the full distribution) with $R>R_c$ and then weighting each event in the counting by $1/R$.
\end{itemize}
\item {\it Sideband in $m$}.  As long as the requirement $R(x|m)>R_c$ does not sculpt a localized feature in $m$, one can estimate the background prediction by performing a fit in the $m$ spectrum from the SB and interpolating to the SR.  This is a standard approach, as discussed in Sec.~\ref{sec:intro}.  

\end{itemize}

Further details about background estimation are presented in Sec.~\ref{sec:back} for the numerical example described in the next section.

\subsection{Comparison with the CWoLa hunting method}

The CWoLa hunting method~\cite{Collins:2018epr,Collins:2019jip} is a recently-proposed model-agnostic sideband method that also uses machine learning and will serve as a benchmark for ANODE.  In the CWoLa hunting approach, the signal sensitivity is achieved by training a classifier to distinguish the SR from the SB.  This classifier will approach the likelihood ratio $R_\text{CWoLa}$, which is optimal under certain conditions:
\beq
\label{eq:rcwola}
R_\text{CWoLa}(x) = {p_\text{data}(x|\text{SR})\over p_\text{data}(x|\text{SB})}= {p_\text{data}(x|\text{SR})\over p_\text{background}(x|\text{SB})}= {p_\text{data}(x|\text{SR})\over p_\text{background}(x|\text{SR})},
\eeq 
where the second equality is true in the absence of signal in the sideband\footnote{This is not strictly necessary - the classifier can still be optimal even if there is some signal in the sideband~\cite{Metodiev:2017vrx}.} and the third equality is true when $x$ and $m$ are independent.  The background is estimated using a sideband fit after placing a selection based on the above classifier.

A key assumption of the CWoLa method is that $x$ and $m$ are independent.  This condition is stronger than the requirement for the background fit, but is necessary for achieving signal sensitivity.   In particular, in the presence of a dependence between $x$ and $m$, the CWoLa classifier will learn the true differences between SB and SR.  If these differences are larger than the difference between signal and background in the SR, the CWoLa classifier may not succeed in finding the signal. 

In contrast, the ANODE method does not require any particular relationship between $x$ and $m$ to achieve signal sensitivity.  In fact, the information about $m$ could be fully contained within $x$, and ANODE could still succeed in principle. Therefore, ANODE can make use of features which are strongly correlated with $m$, thus extending the potential sensitivity to new signals.  This is possible because of the two step density estimation, interpolating $p_\text{background}(x|m)$ from the sideband and then estimating $p_\text{data}(x|m)$ from the SR.   Such an approach is not possible with CWoLa hunting, which directly learns the likelihood ratio.  The only requirement for ANODE is that there are no non-trivial features in the SR that cannot be smoothly predicted from the SB. Section~\ref{sec:correlations} illustrates the ability of ANODE to cope with correlated features.

\section{Details of the Sample}
\label{sec:sim}

A simulated resonance search using large-radius dijets is used to illustrate ANODE.  The simulated datasets are from the LHC Olympics 2020 challenge research and development dataset~\cite{gregor_kasieczka_2019_2629073}.  For a background process, one million quantum chromodynamic (QCD) dijet events are simulated with Pythia~8~\cite{Sjostrand:2006za,Sjostrand:2007gs} without pileup or multiple parton interactions.  The signal is a hypothetical $W'$ boson ($m_{W'}=3.5$~TeV) that decays into an $X$ boson ($m_X=500$~GeV) and a $Y$ boson ($m_Y=100$~GeV), with the same simulation setup as the QCD dijets.  The $X$ and $Y$ bosons decay promptly into quarks and due to their large Lorentz boost in the lab frame, the resulting hadronic decay products are captured by a single large-radius jet.  The detector simulation is performed with Delphes~3.4.1~\cite{deFavereau:2013fsa,Mertens:2015kba,Selvaggi:2014mya} and particle flow objects are clustered into jets using the Fastjet~\cite{Cacciari:2011ma,Cacciari:2005hq} implementation of the anti-$k_t$ algorithm~\cite{Cacciari:2008gp} using $R=1.0$ as the jet radius.  Events are selected by requiring at least one such jet with $p_T>1.3$ TeV.  While there exist LHC searches for the case that $X$ and $Y$ are electroweak bosons~\cite{Aad:2019fbh,Sirunyan:2019jbg}, the generic case is currently uncovered by a dedicated search.

The resonant feature $m$ will be the invariant mass of the leading two jets, $m_{JJ}$.  These two jets are ordered by their mass $m_J$ so that by construction, $m_{J_1}<m_{J_2}$.  The discriminating features $x$ are four-dimensional, consisting of the observables:
\beq
\label{eq:x}
m_{J_1},\,\,\, m_{J_2}-m_{J_1}, \,\,\, \tau_{21}^{J_1},\,\,\, \tau_{21}^{J_2},
\eeq
where $\tau_{21}$ is the n-subjettiness ratio~\cite{Thaler:2011gf,Thaler:2010tr}.  This observable is the most widely used single feature for identifying jets with a two-prong substructure.
While the ultimate goal of ANODE is to perform density estimation on high-dimensional, low-level features, there is already utility in a search with high-level features from Eq.~\ref{eq:x}. Thus to demonstrate how ANODE works, this will be the focus for the rest of this paper.

\begin{figure}[h!]
\centering
\includegraphics[scale=0.55]{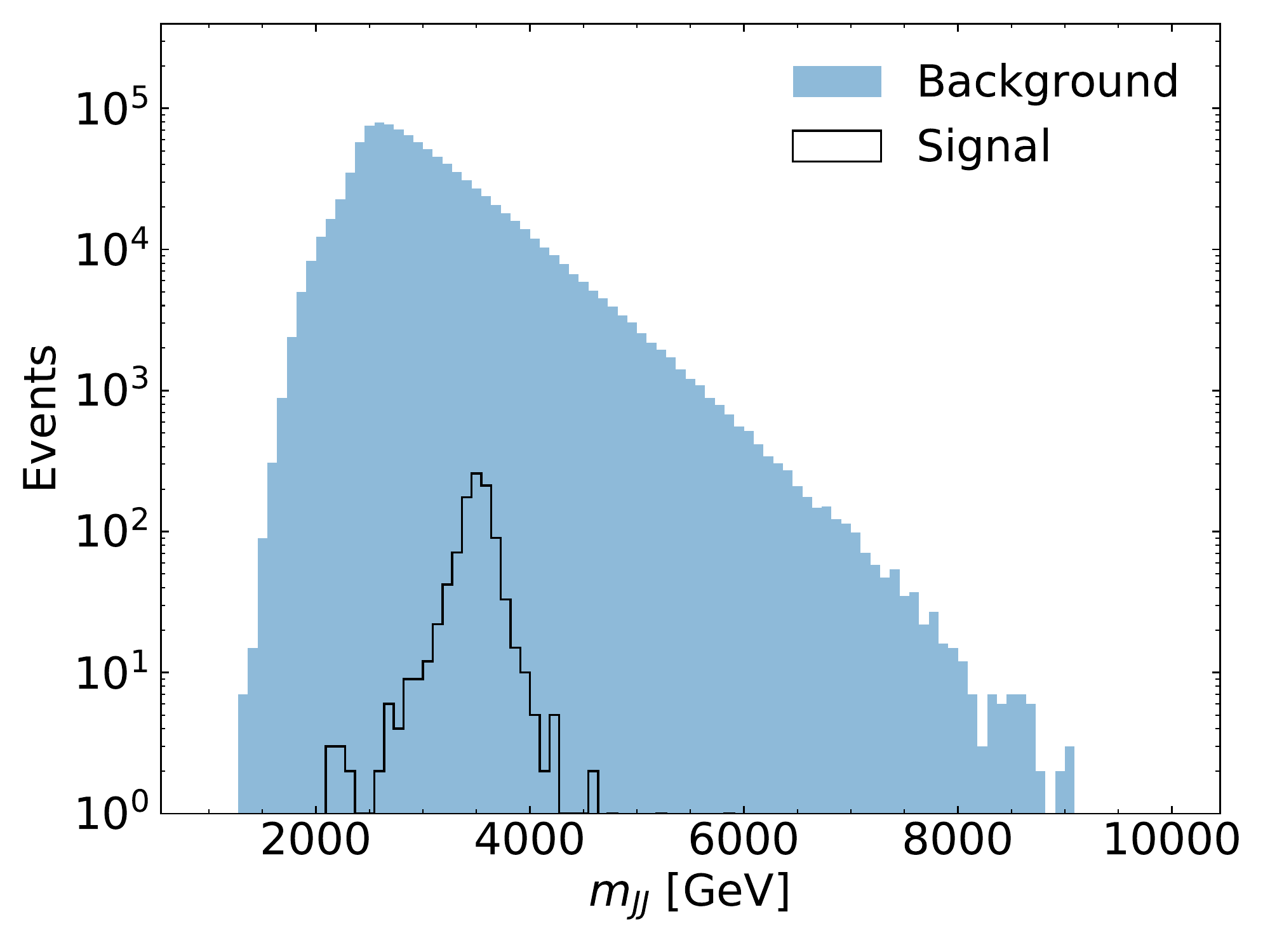}
\caption{Histograms for the invariant mass of the leading two jets for the Standard Model background as well as the injected signal.  There are 1 million background events and 1000 signal events.}
\label{fig:mJJlhco}
\end{figure}

Simulated data are constructed by injecting 1000 signal events to the full background sample.  A histogram of $m_{JJ}$ is presented in Fig.~\ref{fig:mJJlhco}.  As expected, the signal peaks near $m_{W'}$.  The signal region is defined by $m_{JJ}\in[3.3,3.7]$~TeV and then the sideband is the rest of the spectrum.  The simulated data are divided into two equal samples for training and testing; thus we have $\approx 500,000$ background and $\approx 500$ signal events in each sample. In the SR, we are left with $\approx 60,000$ background and $\approx 400$ signal events in each sample. This corresponds to $S/\sqrt{B}=1.6$ and $S/B=0.6\%$ in the SR.  This value of $S/\sqrt{B}$ would be the approximate significance from a sideband fit (ignoring the fit errors).  Section~\ref{sec:sens} will show how much this can be enhanced from ANODE.

\begin{figure}[t!]
 \centering
\includegraphics[scale=0.4]{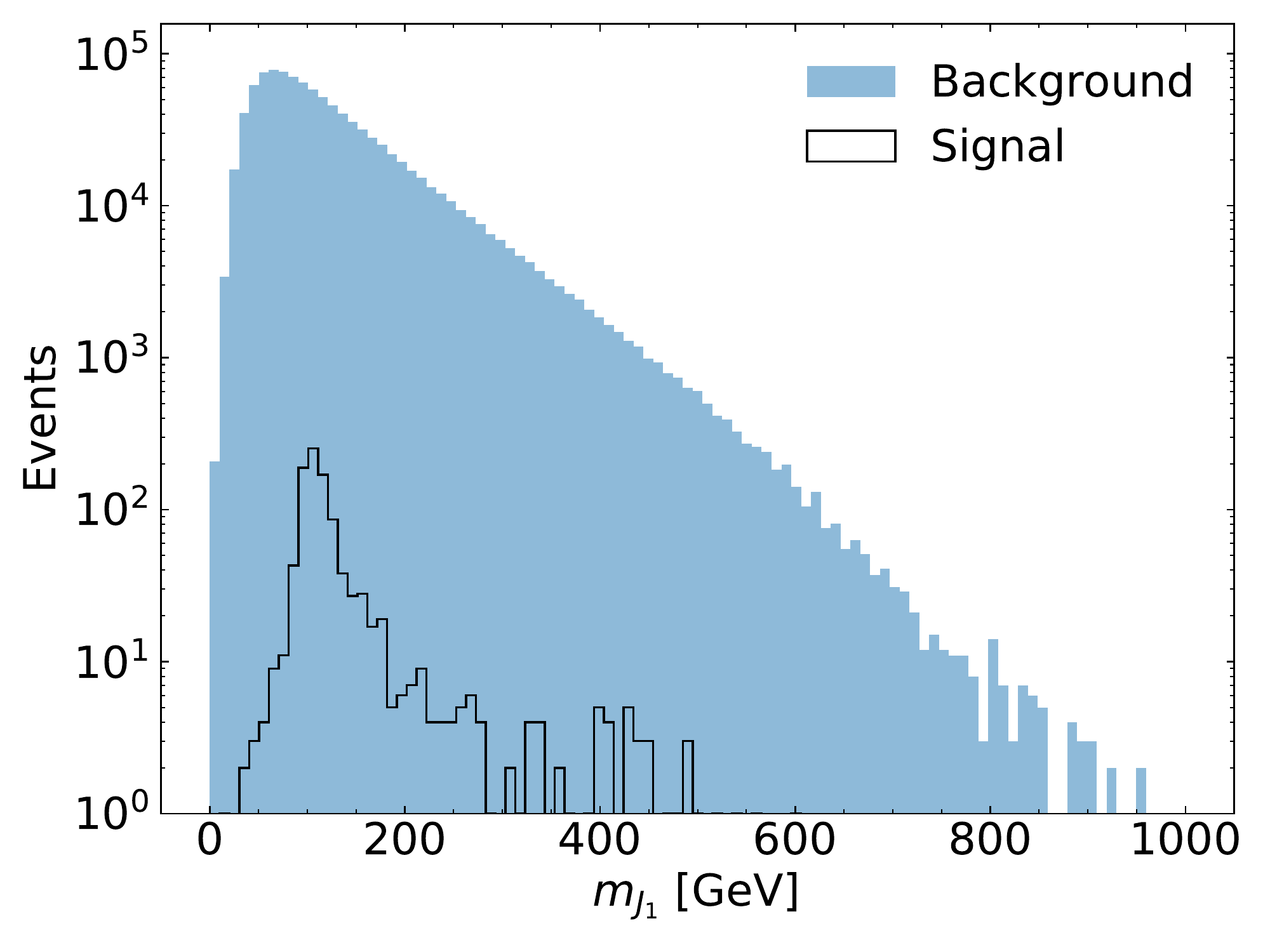}\includegraphics[scale=0.4]{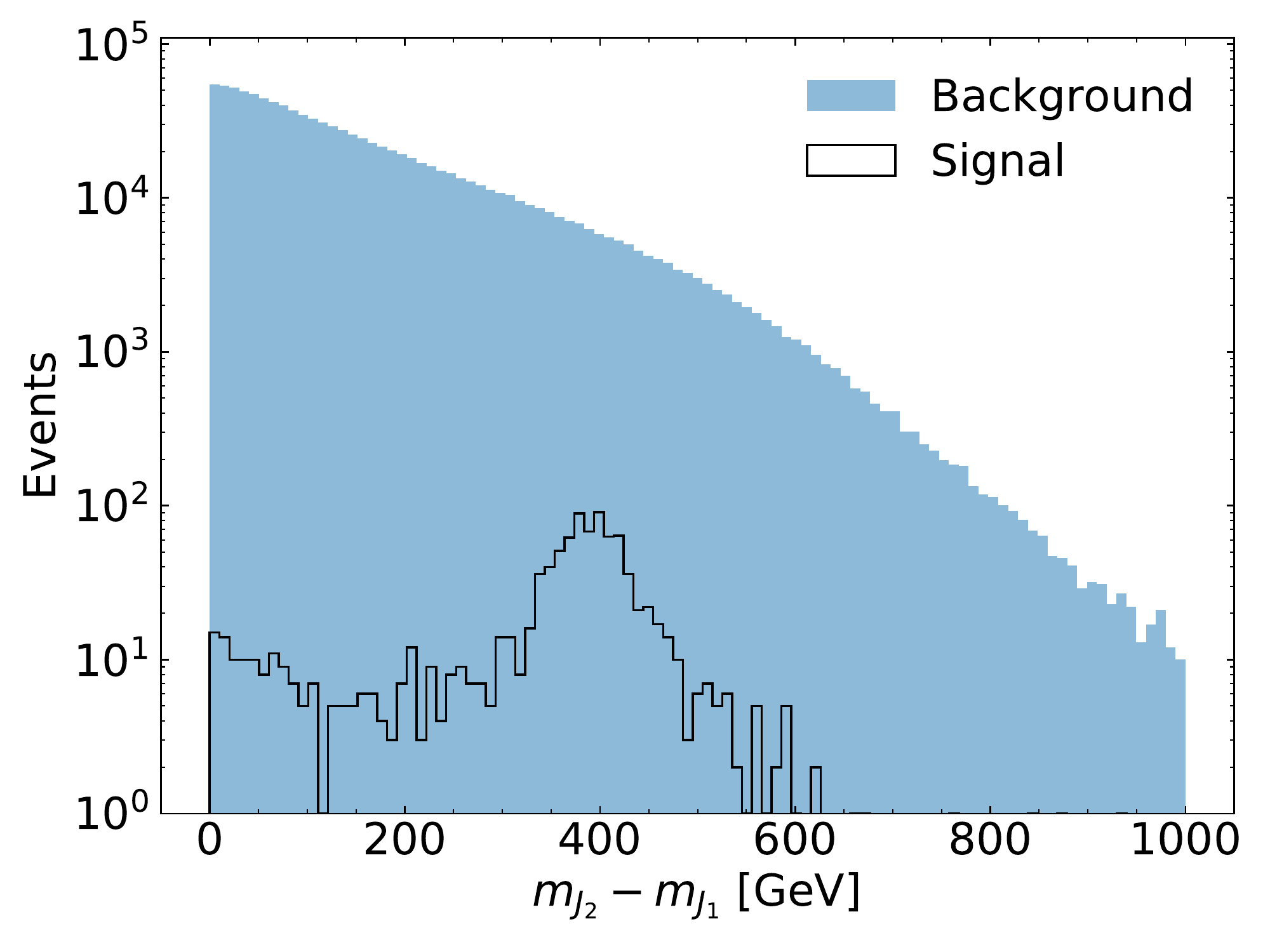}\\
\includegraphics[scale=0.4]{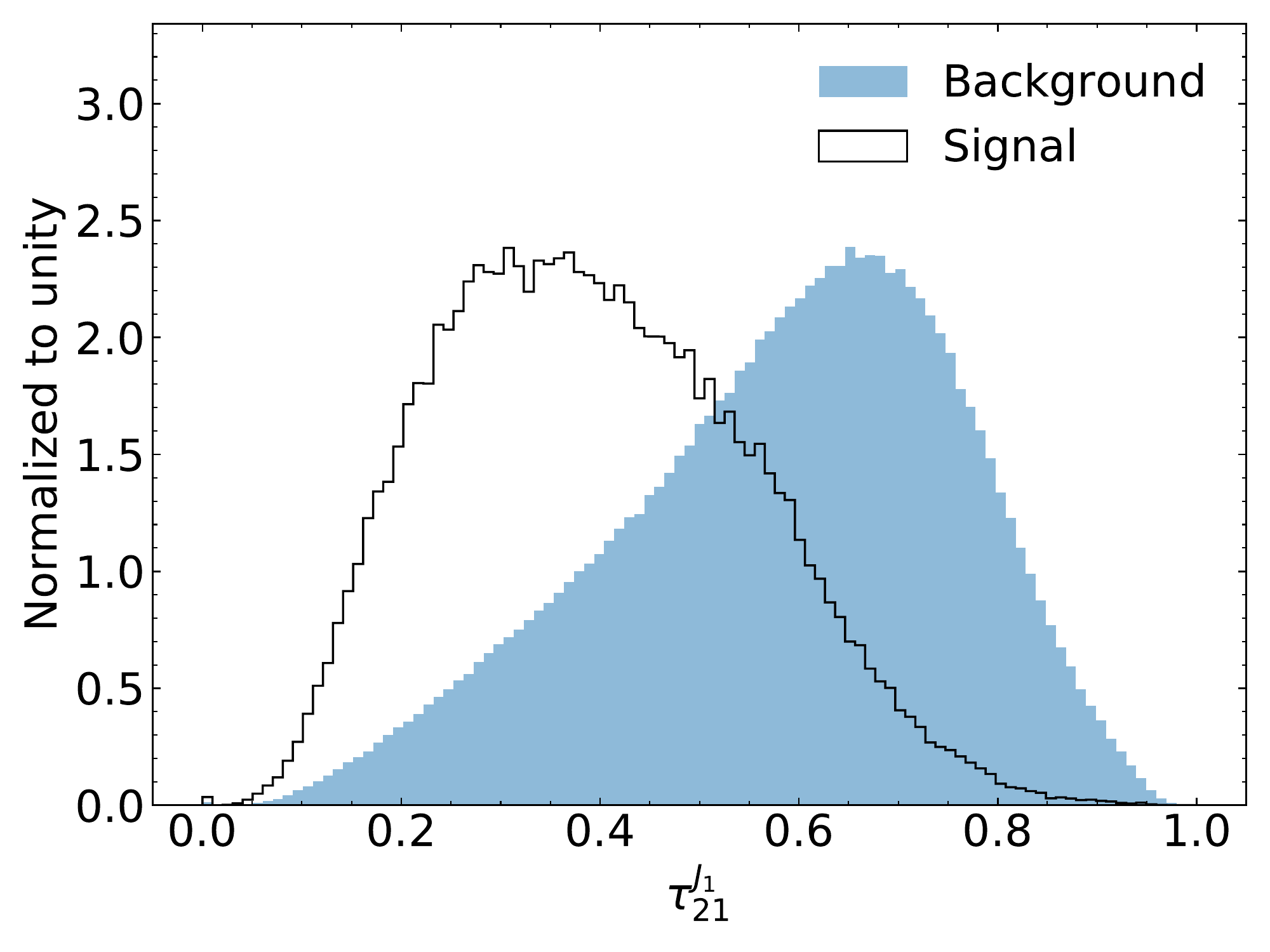}\includegraphics[scale=0.4]{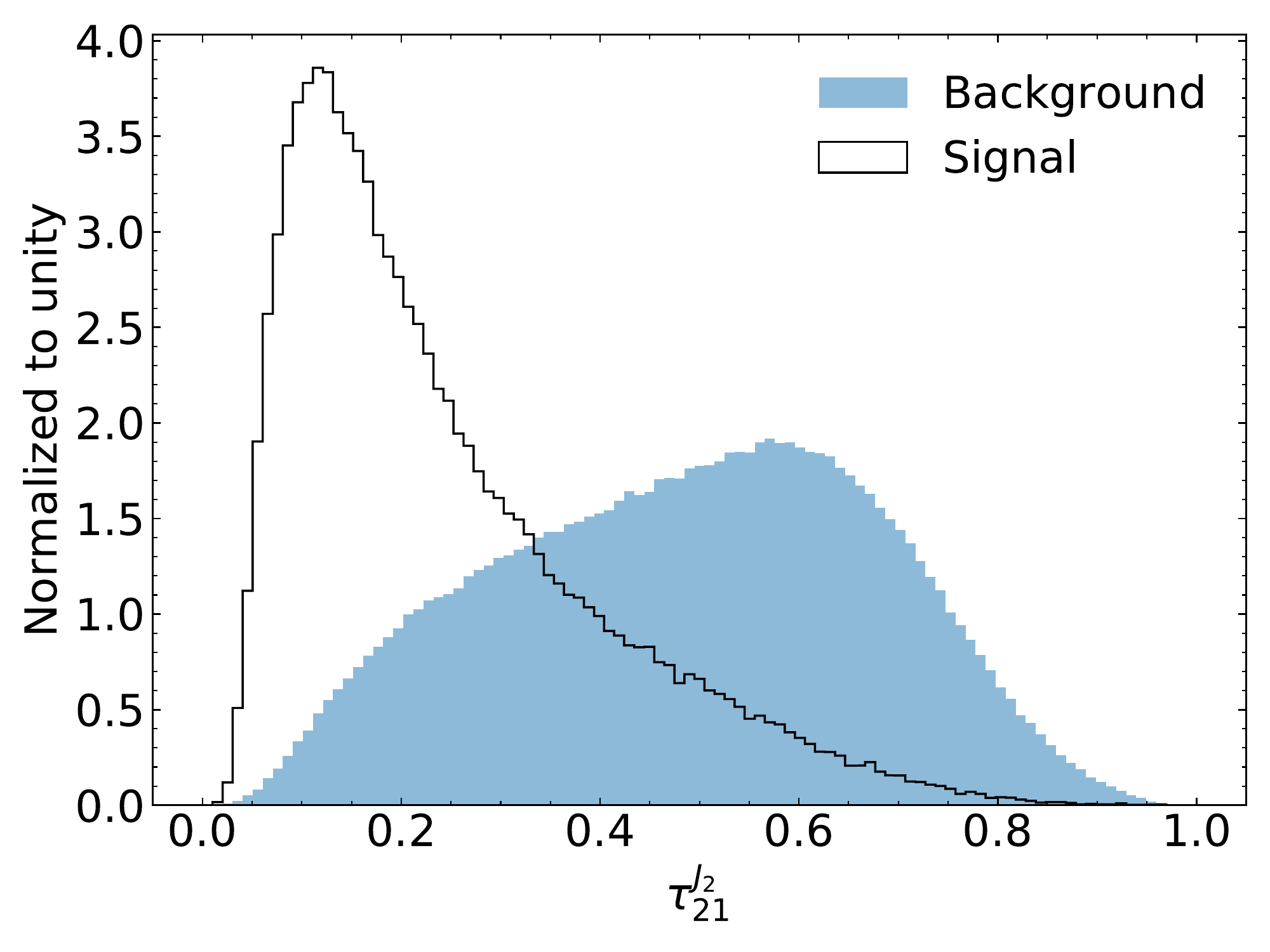}
\caption{The four features used for classification: $m_{J_1}$ (top left), $m_{J_1}-m_{J_2}$ (top right), $\tau_{21}^{J_1}$ (bottom left), and $\tau_{21}^{J_2}$ (bottom right).  These histograms are inclusive in $m_{JJ}$.  There are 1 million background events and 1000 signal events for the mass histograms. }
\label{fig:xlhco}
\end{figure}

The additional four features for classification are shown in Fig.~\ref{fig:xlhco}.  The lighter jet mass peaks near $m_Y$ and the difference between masses peaks at about $m_X-m_Y=400$ GeV.   The $\tau_{21}$ observables are lower for the two-prong signal jets than for the mostly one-prong background jets.  Jet mass and $\tau_{21}$ are negatively correlated for QCD jets~\cite{Dolen:2016kst} and so $\tau_{21}$ is higher for $J_2$ than for $J_1$.  

The conditional MAF (along with most methods of density estimation) has difficulty at sharp, discontinuous edges and boundaries, so we first transform the dataset before performing density estimation.  First, all features are linearly scaled to be $(\text{feature})\mapsto x\in[0,1]$.  Then, the logit transformation $\log(x/(1-x))$ is applied to map the scaled features to be between $(-\infty,\infty)$.  The Jacobian for this map is accounted for when computing probability densities for the original feature space.  Even with this transformation, density estimation is difficult near the boundaries.  Therefore, the scaled features are required to have $0.05<x<0.95$.  This keeps 95\% (72\%) of the signal (background) in the SR.  Below we will refer to this as the ``fiducial region.'' All results below are computed with respect to the number of events after this truncation. 

\section{Results}
\label{sec:results}

\subsection{Sensitivity}
\label{sec:sens}

The conditional MAF is optimized\footnote{Based on code from \url{https://github.com/ikostrikov/pytorch-flows}.} using the log likelihood loss function, $\log(p(x|m))$.  All of the neural networks are written in PyTorch~\cite{NEURIPS2019_9015}.  For the hyperparameters, there are 15 MADE blocks (one layer each) with 128 hidden units per block.  Networks are optimized with Adam~\cite{adam} using a learning rate $10^{-4}$ and weight decay of $10^{-6}$.  The SR and SB density estimators are each trained for 50 epochs. No systematic attempt was made to optimize these hyperparameters and it is likely that better performance could be obtained with further optimization. For the SR density estimator, the last epoch is chosen for simplicity and it was verified that the results are robust against this choice.  The SB density estimator significantly varies from epoch to epoch.  Averaging the density estimates point-wise over 10 consecutive epochs results in a stable result.  Averaging over more epochs does not further improve the stability.  All results with ANODE present the SB density estimator with this averaging scheme for the last 10 epochs.

\begin{figure}[t!]
\centering
\includegraphics[scale=0.55]{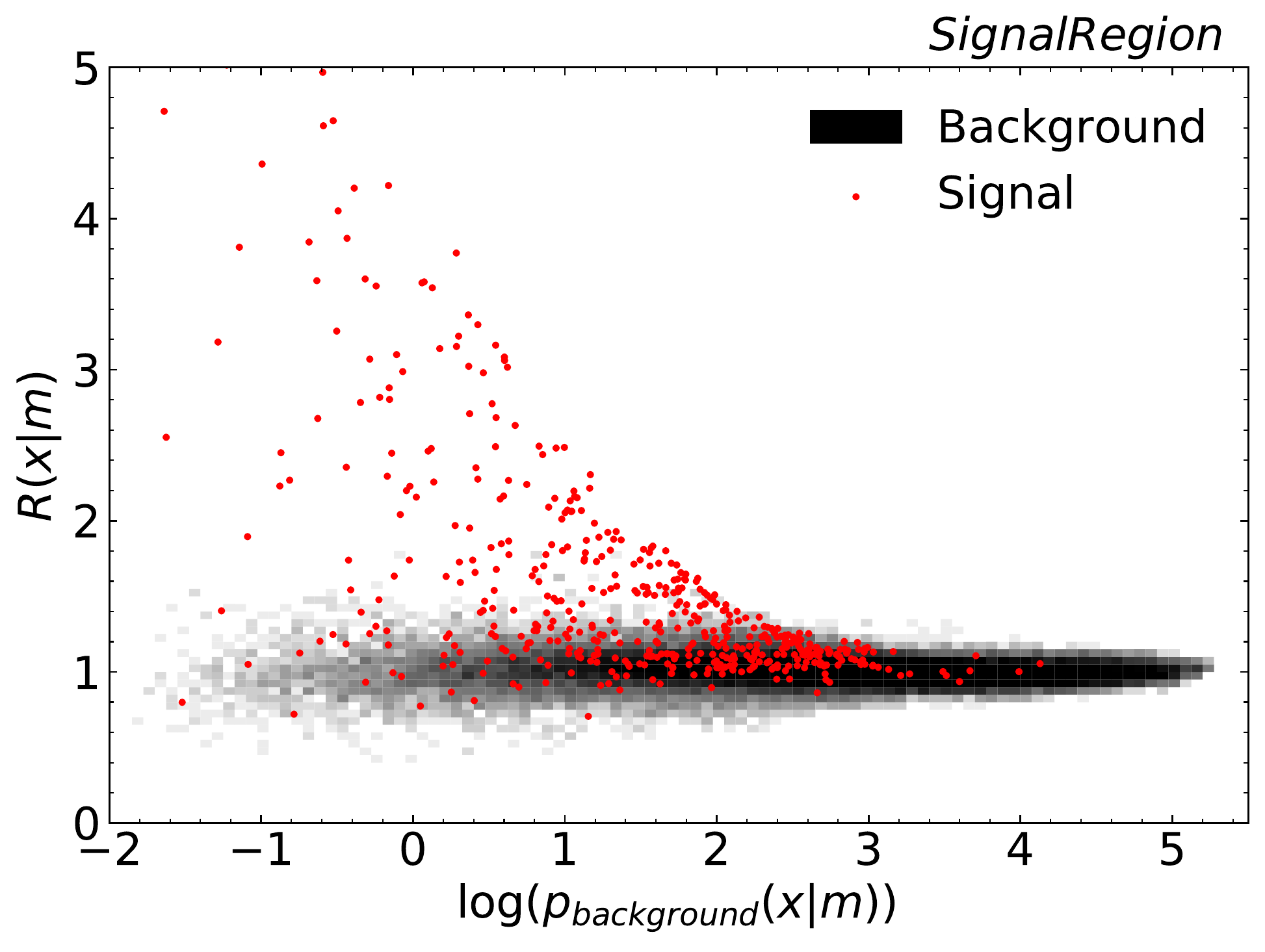}
\caption{Scatter plot of $R(x|m)$ versus $\log p_\text{background}(x|m)$ across the test set in the SR.  Background events are shown (as a two-dimensional histogram) in grayscale and individual signal events are shown in red.}
\label{fig:RvslogPbg}
\end{figure}

Figure~\ref{fig:RvslogPbg} shows a scatter plot of $R(x|m)$ versus $\log p_\text{background}(x|m)$ for the test set in the SR.  As desired, the background is mostly concentrated around $R(x|m)=1$, while there is a long tail for signal events at higher values of $R(x|m)$ and between $-2<\log p_\text{background}(x|m) <2$.   This is exactly what is expected for this signal: it is an over-density ($R>1$) in a region of phase space that is relatively rare for the background ($p_\text{background}(x|m)\ll 1$).

The background density in Fig.~\ref{fig:RvslogPbg} also shows that the $R(x|m)$ is narrower around $1$ when $p_\text{background}(x|m)$ is large and more spread out when $p_\text{background}(x|m)\ll 1$.    This is evidence that the density estimation is more accurate when the densities are high and worse when the densities are low. This is also to be expected: if there are many data points close to one another, it should be easier to estimate their density than if the data points are very sparse. 

\begin{figure}[t!]
\centering
\includegraphics[scale=0.35]{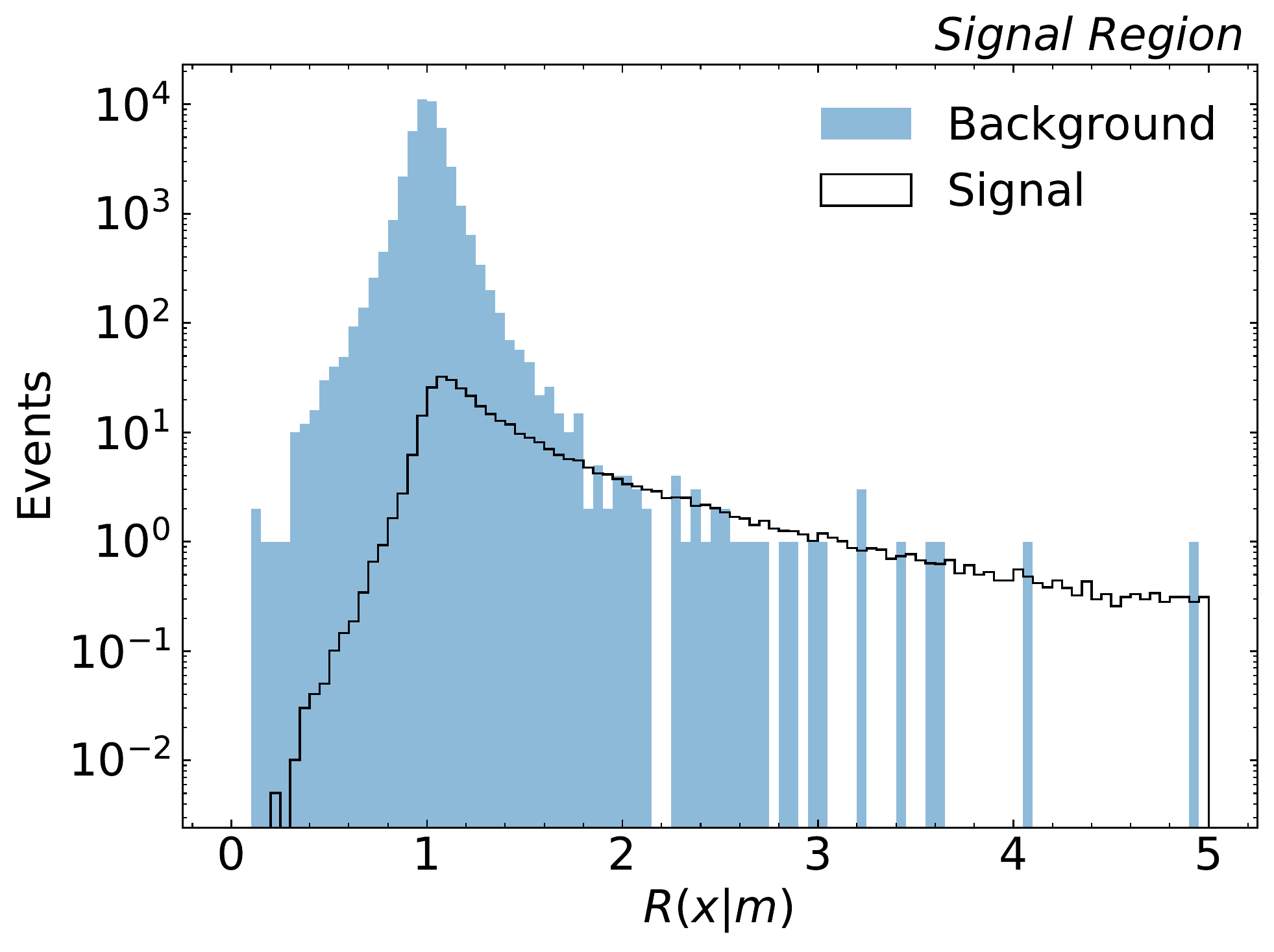}\hspace{5mm}\includegraphics[scale=0.35]{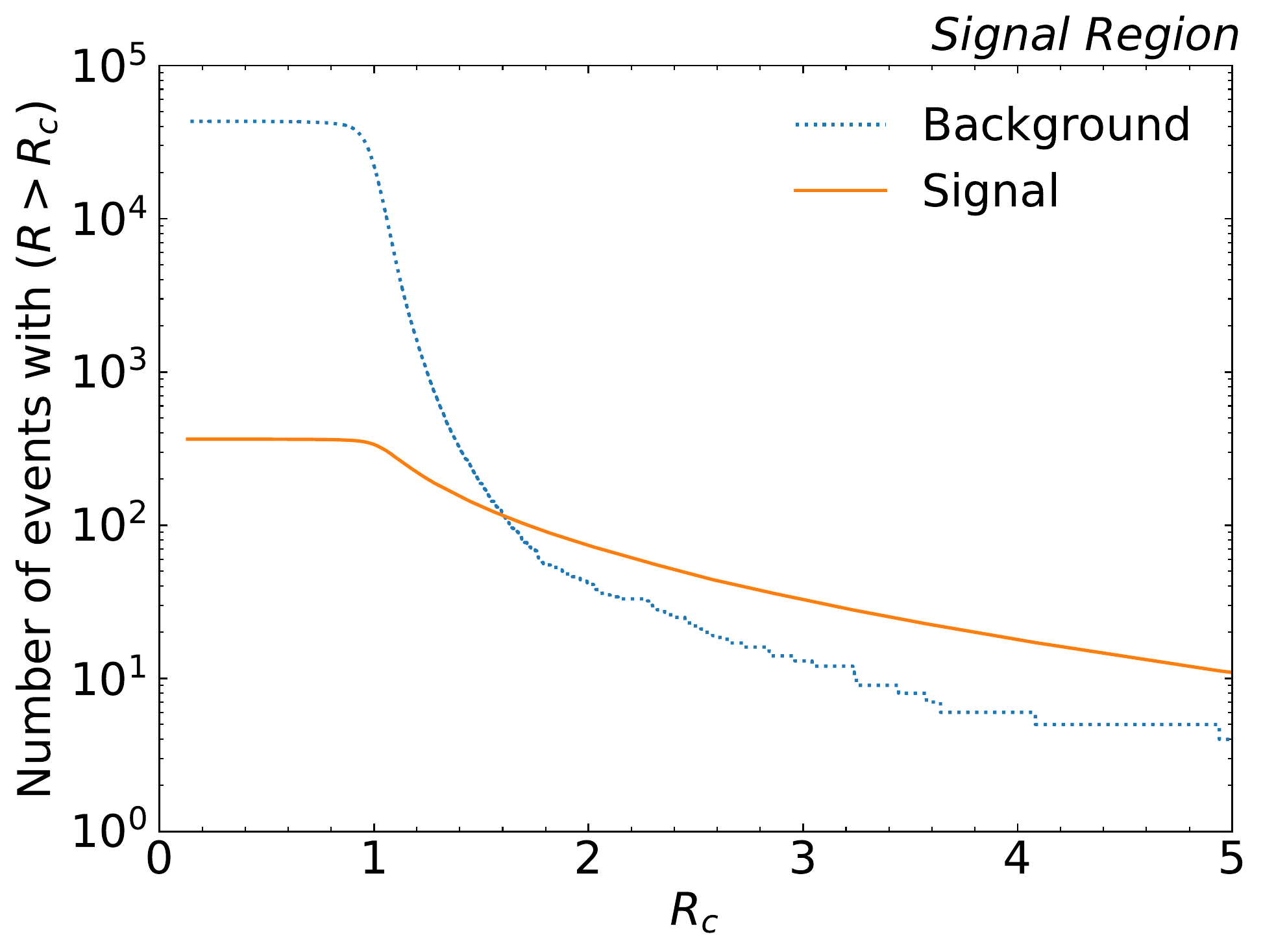}
\caption{Left: Histogram of $R(x|m)$ evaluated on the test set; Right: the integrated number of events that survive a threshold on $R(x|m)$.  The two distributions are scaled to represent the rates for 500,000 total background events and 500 total signal events, as introduced in Sec.~\ref{sec:sim}.}
\label{fig:Rhist}
\end{figure}

\begin{figure}[t!]
\centering
\includegraphics[scale=0.35]{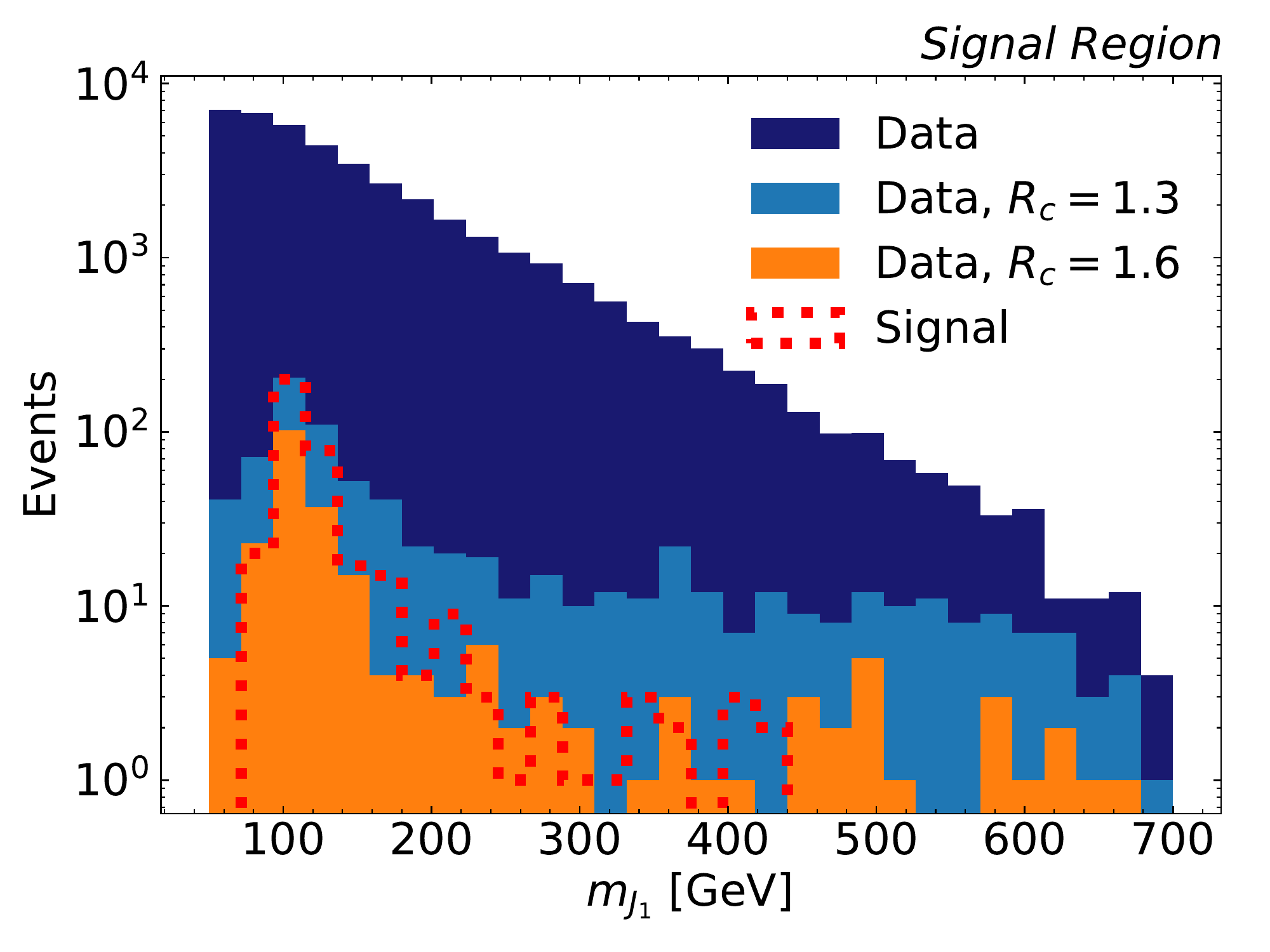}\hspace{5mm}\includegraphics[scale=0.35]{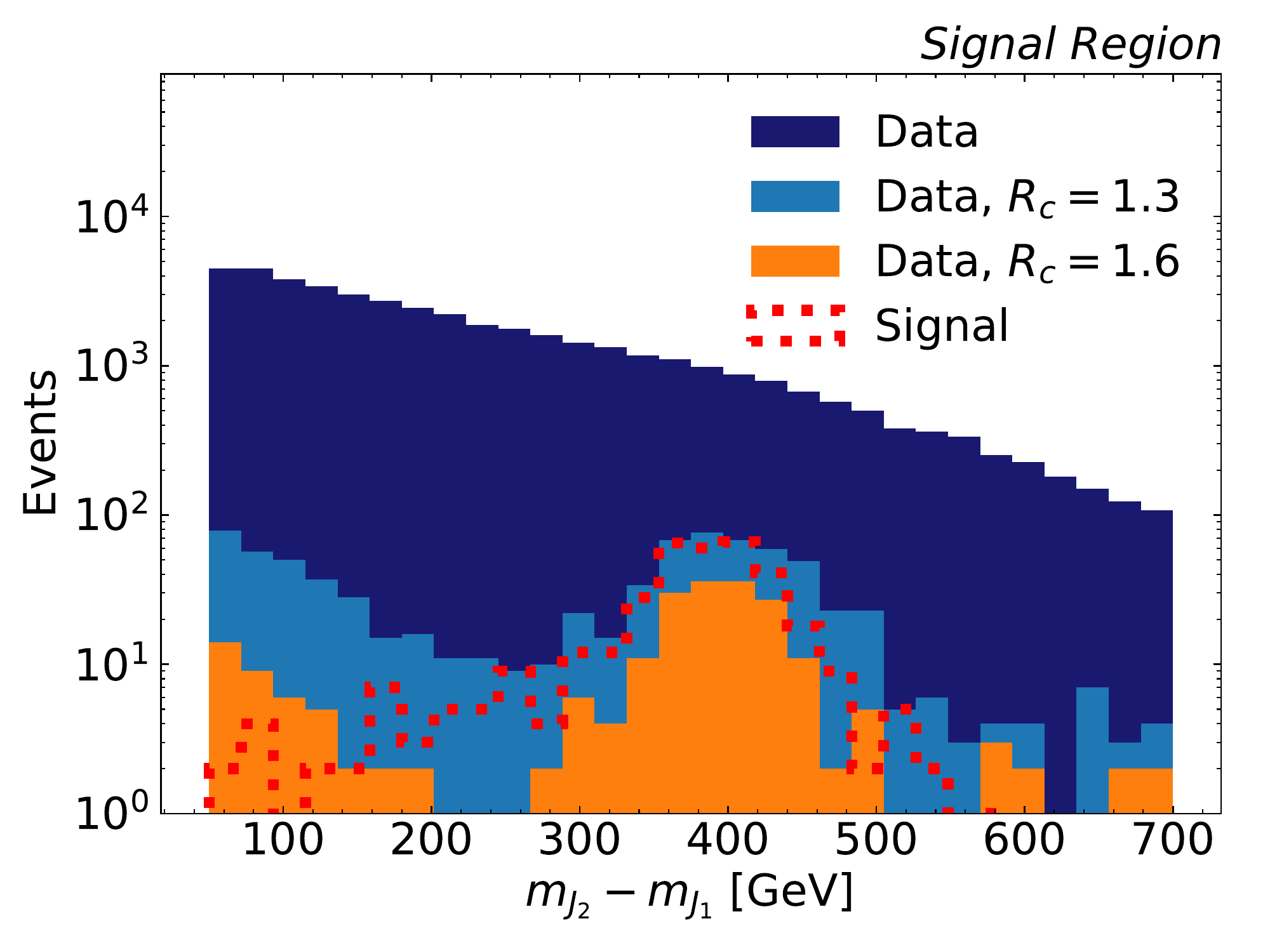}
\caption{Distributions of $m_{J_1}$ (left) and $m_{J_2}-m_{J_1}$ (right) in the signal region after applying a threshold requirement on $R$.}
\label{fig:aftercuts}
\end{figure}

Another view of the results is presented in Fig.~\ref{fig:Rhist}, with one-dimensional information about $R(x|m)$ in the SR.  The left plot of Fig.~\ref{fig:Rhist} shows that the background is centered and approximately symmetric around $R=1$ with a standard deviation of approximately 17\%.  This width is due to various sources, including the accuracy of the SR density, the accuracy of the SB density, and the quality of the interpolation from SB to SR.  Each of these sources has contributions from the finite size of the datasets used for training, the neural network flexibility, and the training procedure.  The right plot of Fig.~\ref{fig:Rhist} presents the number of background and signal events as a function of a threshold $R>R_c$.  The starting point are the original numbers background (40,000) and signal (400) numbers in the SR window and the fiducial window.  Starting from low $S/B$ and $S/\sqrt{B}$ one can achieve $S/B> 1$ and a high $S/\sqrt{B}$ with a threshold requirement on $R$.  Figure~\ref{fig:aftercuts} shows that the signal is clearly visible in the $x$ distribution after applying such a threshold requirement.

\begin{figure}[t!]
\centering
\includegraphics[scale=0.38]{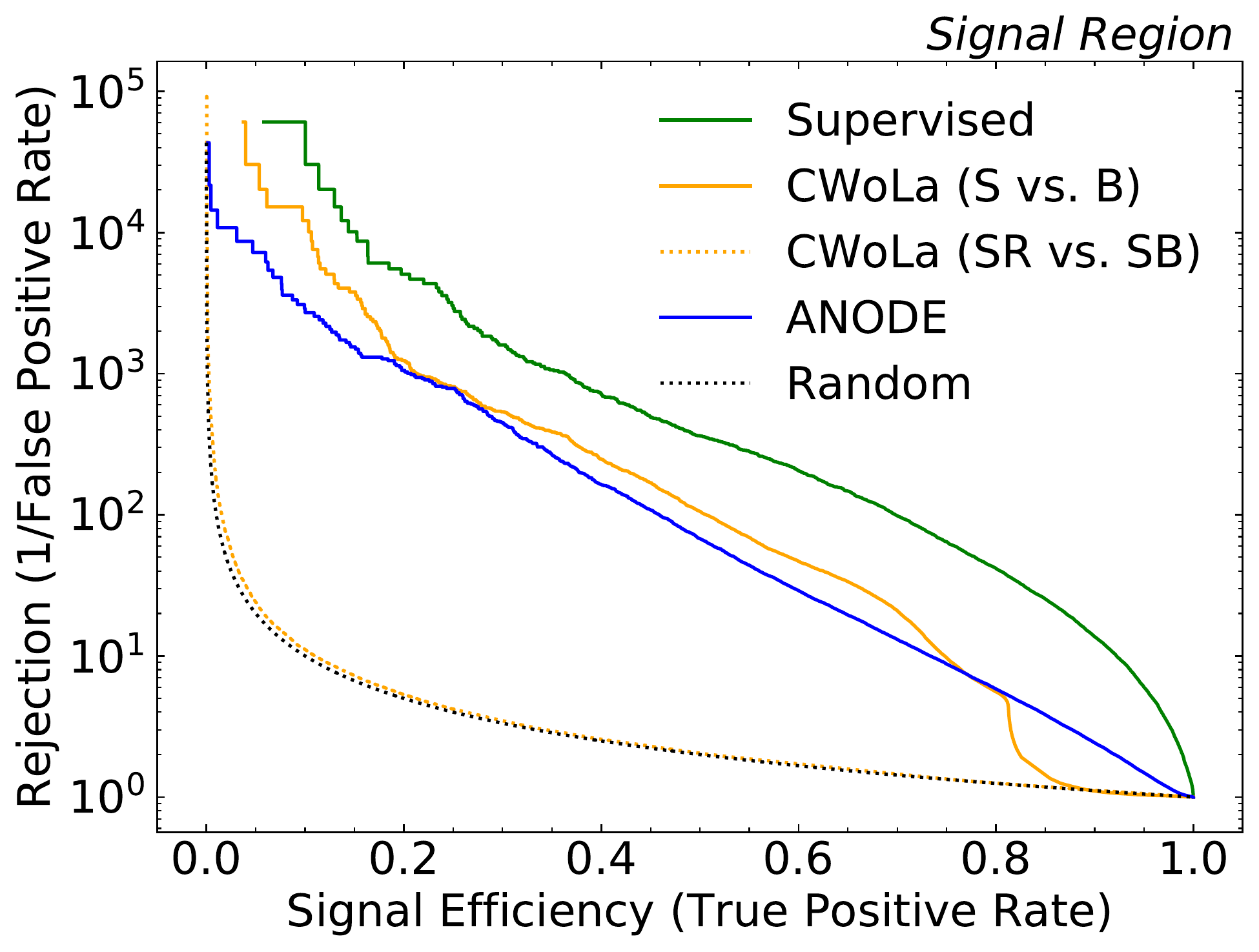}\hspace{5mm}\includegraphics[scale=0.38]{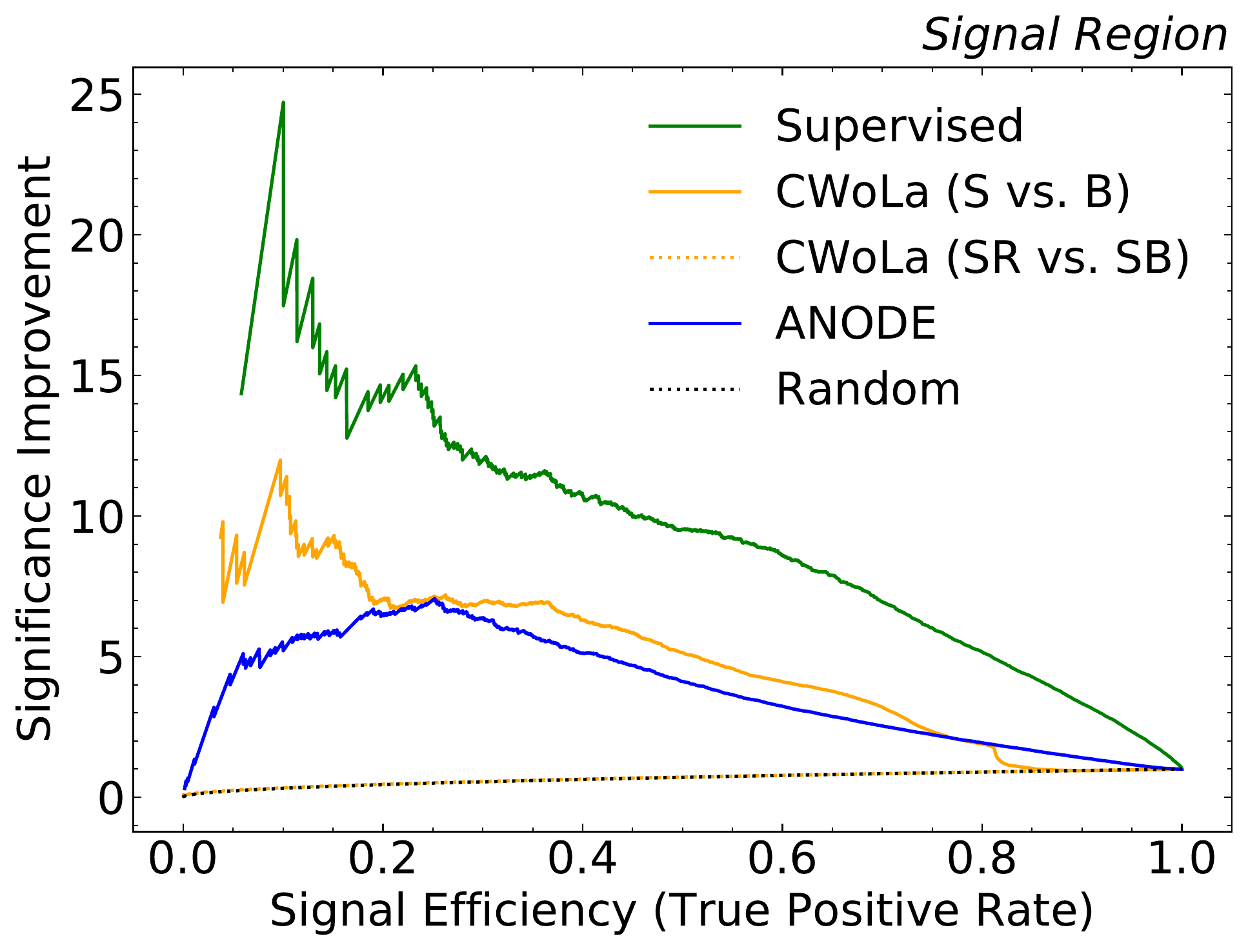}
\caption{Receiver Operating Characteristic (ROC) curve (left) and Significance Improvement Characteristic (SIC) curve (right).  }
\label{fig:ROC}
\end{figure}

The performance of $R$ as an anomaly detector is further quantified by the Receiver Operating Characteristic (ROC) and Significance Improvement Characteristic (SIC) curves in Fig.~\ref{fig:ROC}.  These metrics are obtained by scanning $R$ and computing the signal efficiency (true positive rate) and background efficiency (false positive rate) after a threshold requirement on $R$.  The Area Under the Curve (AUC) for ANODE is 0.82. For comparison, the CWoLa hunting approach is also shown in the same plots.  The CWoLa classifier is trained using sideband regions that are 200 GeV wide on either side of the SR.  The sidebands are weighted to have the same number of events as each other and in total, the same as the SR.  A single NN with four hidden layers with 64 nodes each is trained using Keras~\cite{keras} and TensorFlow~\cite{tensorflow}.   Dropout~\cite{JMLR:v15:srivastava14a} of 10\% is used for each intermediate layer.  Intermediate layers use rectified linear unit activation functions and the last layer uses a sigmoid.  The classifier is optimized using binary cross entropy and is trained for 300 epochs.  As with ANODE, 10 epochs are averaged for the reported results\footnote{A different regularization procedure was used in Ref.~\cite{Collins:2018epr,Collins:2019jip} based on the validation loss and $k$-folding.  The averaging here is expected to serve a similar purpose.}.

The performance of ANODE is comparable to CWoLa hunting in Fig.~\ref{fig:ROC}, which does slightly better at higher signal efficiencies and much better at lower signal efficiencies.  This may be a reflection of the fact that CWoLa makes use of supervised learning and directly approaches the likelihood ratio, while ANODE is unsupervised and attempts to learn both the numerator and denominator of the likelihood ratio.  With this dataset, ANODE is able to enhance the signal significance by about a factor of 7 and would therefore be able to achieve a local significance above $5\sigma$ given that the starting value of $S/\sqrt{B}$ is 1.6.

\subsection{Background Estimation}
\label{sec:back}

\begin{figure}[h!]
\centering
\includegraphics[scale=0.35]{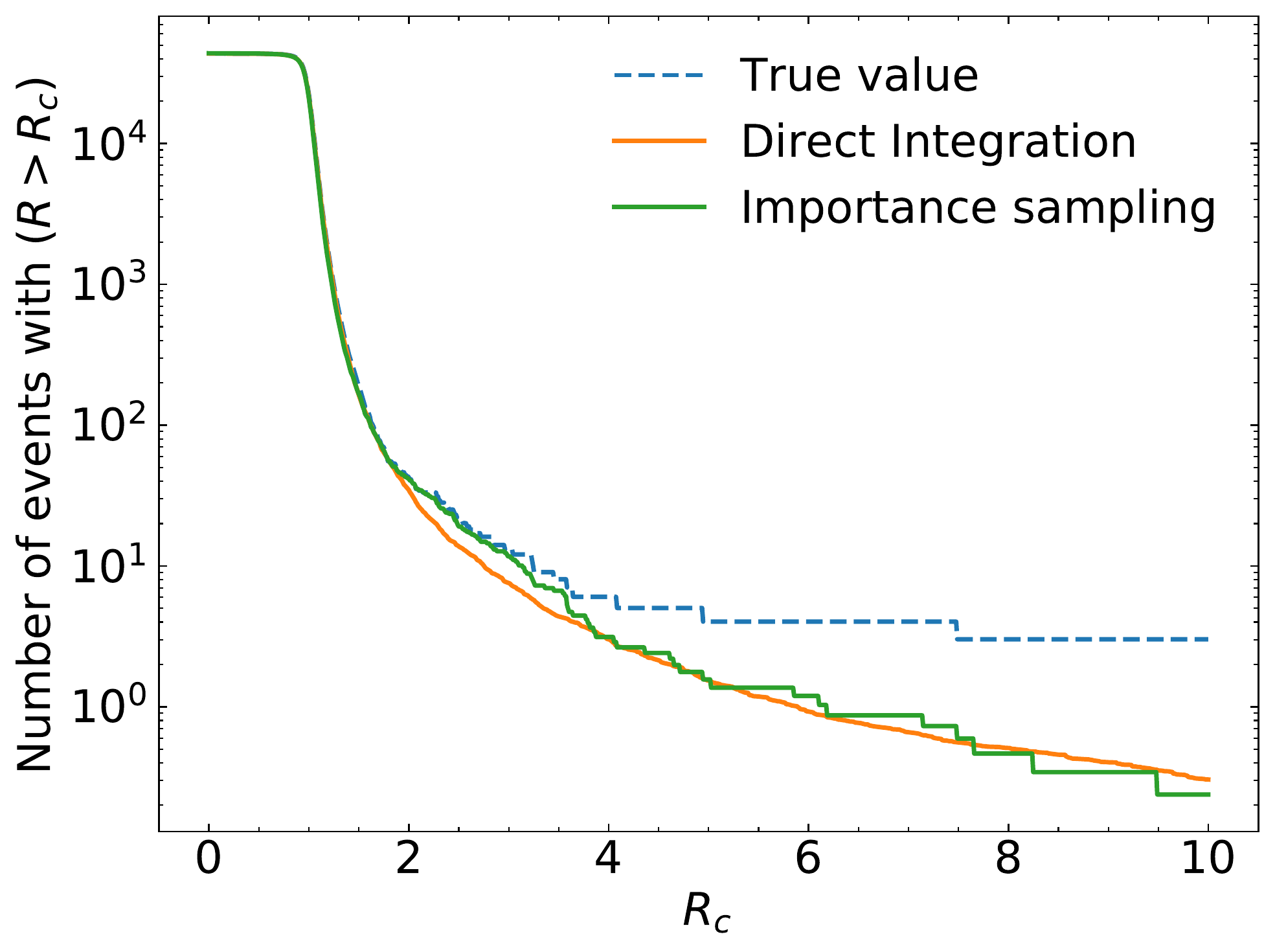}\hspace{5mm}\includegraphics[scale=0.36]{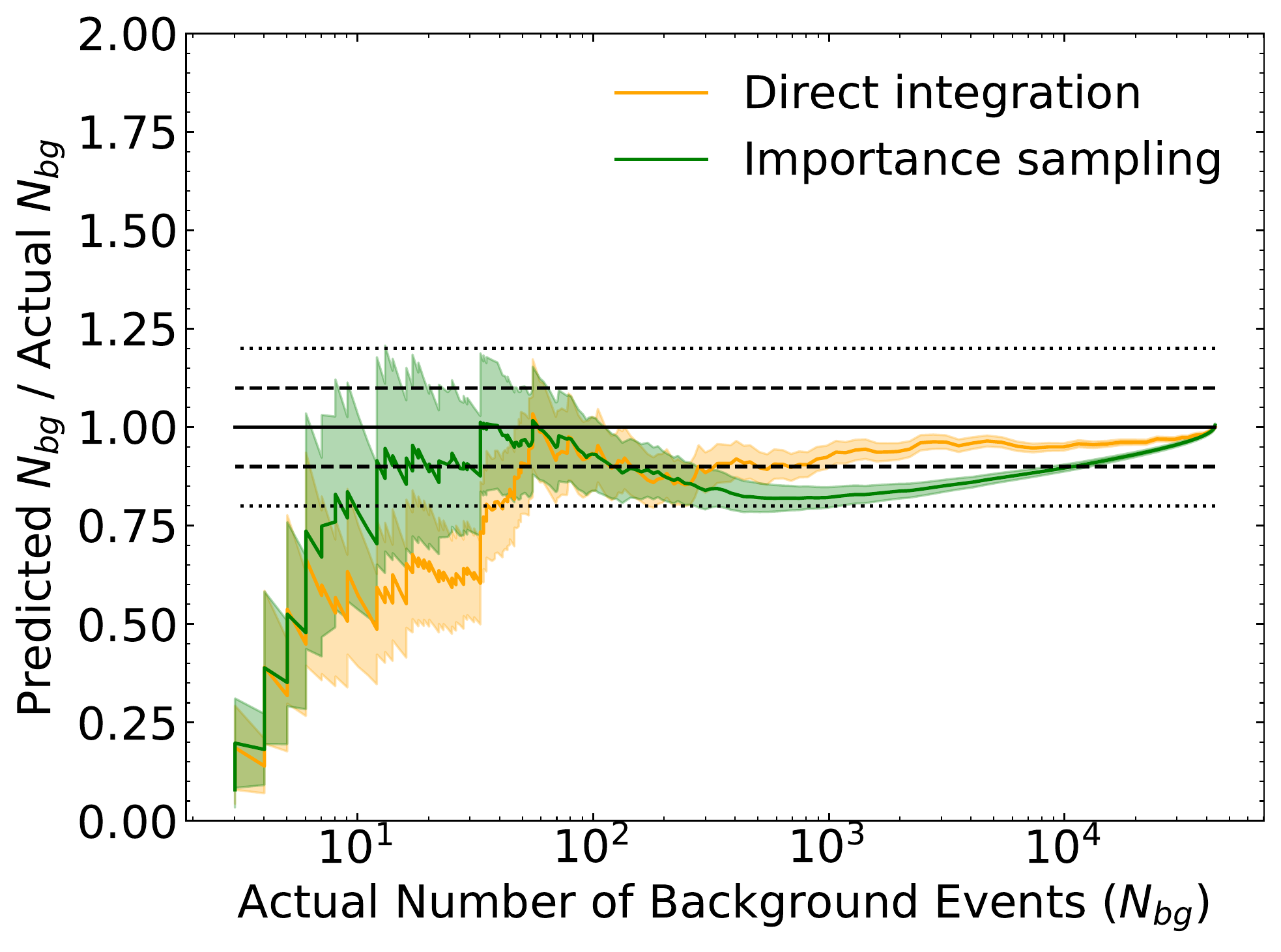}
\caption{Left: The number of events after a threshold requirement $R>R_c$ using the two integration methods described in Sec.~\ref{sec:methodbackground}, as well as the true background yield.  Right: The ratio of the predicted and true background yields from the left plot, as a function of the actual number of events that survive the threshold requirement.  The shaded bands around the central predictions are the $1\sigma$ statistical (Poisson) uncertainty derived from the observed background counts. The black dashed and dotted lines are 10\% and 20\% around a ratio of $1$.}
\label{fig:Nbgpredratio_noshift}
\end{figure}

This section explores the possibility of using the estimate of $p_\text{background}(x|m)$ to directly determine the background efficiency in the SR after a requirement on $R>R_c$.  Figure~\ref{fig:Nbgpredratio_noshift} presents a comparison between integration methods (direct integration and importance sampling) described in Sec.~\ref{sec:methodbackground} and the true background yields.  Qualitatively, both methods are able to characterize the yield across several orders of magnitude in background efficiency.  However, both methods diverge from the truth in the extreme tails of the $R$ distribution.  The right plot of Fig.~\ref{fig:Nbgpredratio_noshift} offers a quantitative comparison between methods.  For efficiencies down to about $10^{-3}$, both methods are accurate within about 25\%.  The direct integration method has a smaller bias of about 10\%.   This is consistent with Fig.~\ref{fig:Rhist}, for which the standard deviation is between 10-20\%.

\subsection{Performance on a Dataset with Correlated Features}
\label{sec:correlations}

\begin{figure}[h!]
\centering
\includegraphics[scale=0.35]{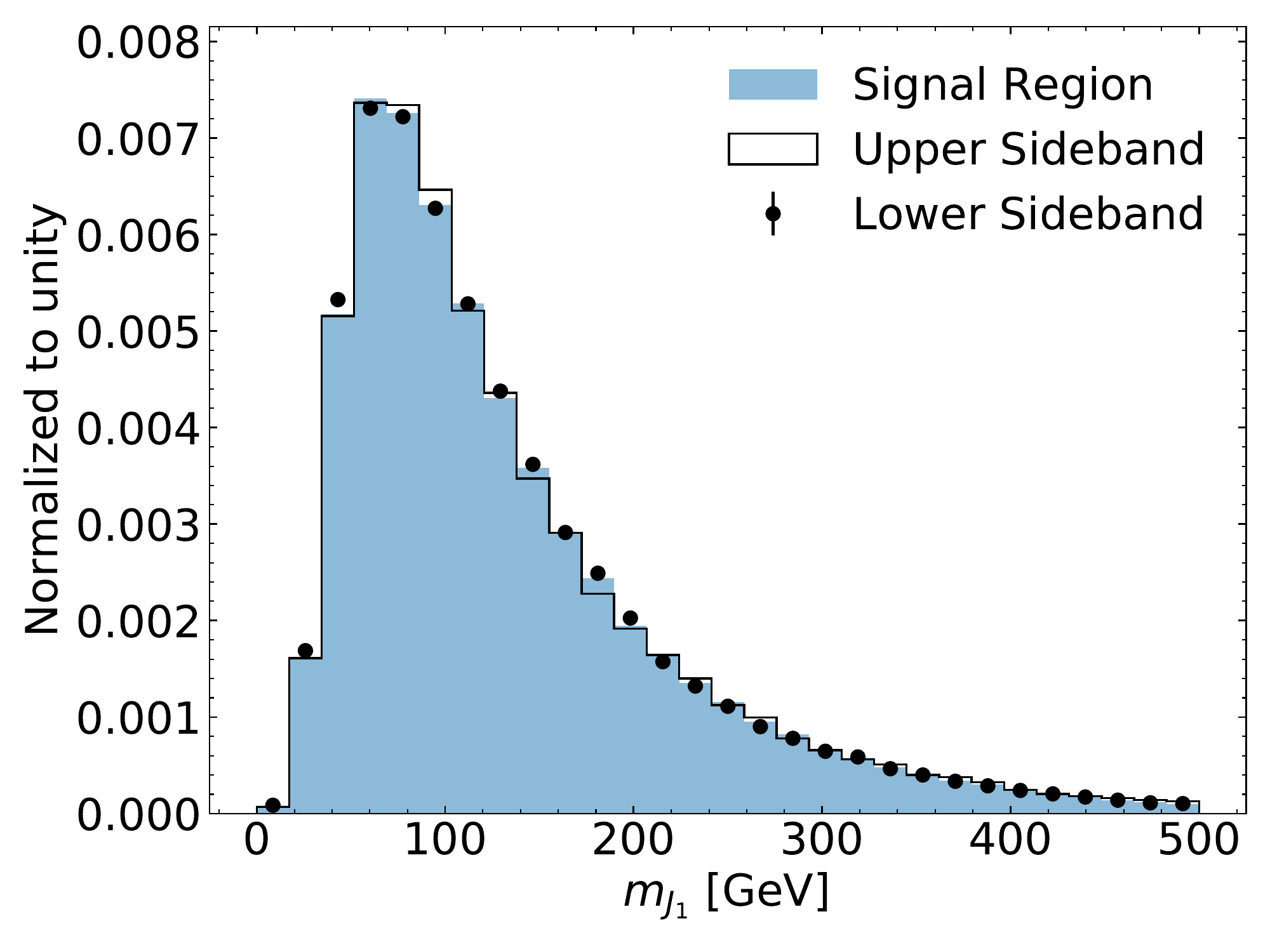}\hspace{5mm}\includegraphics[scale=0.35]{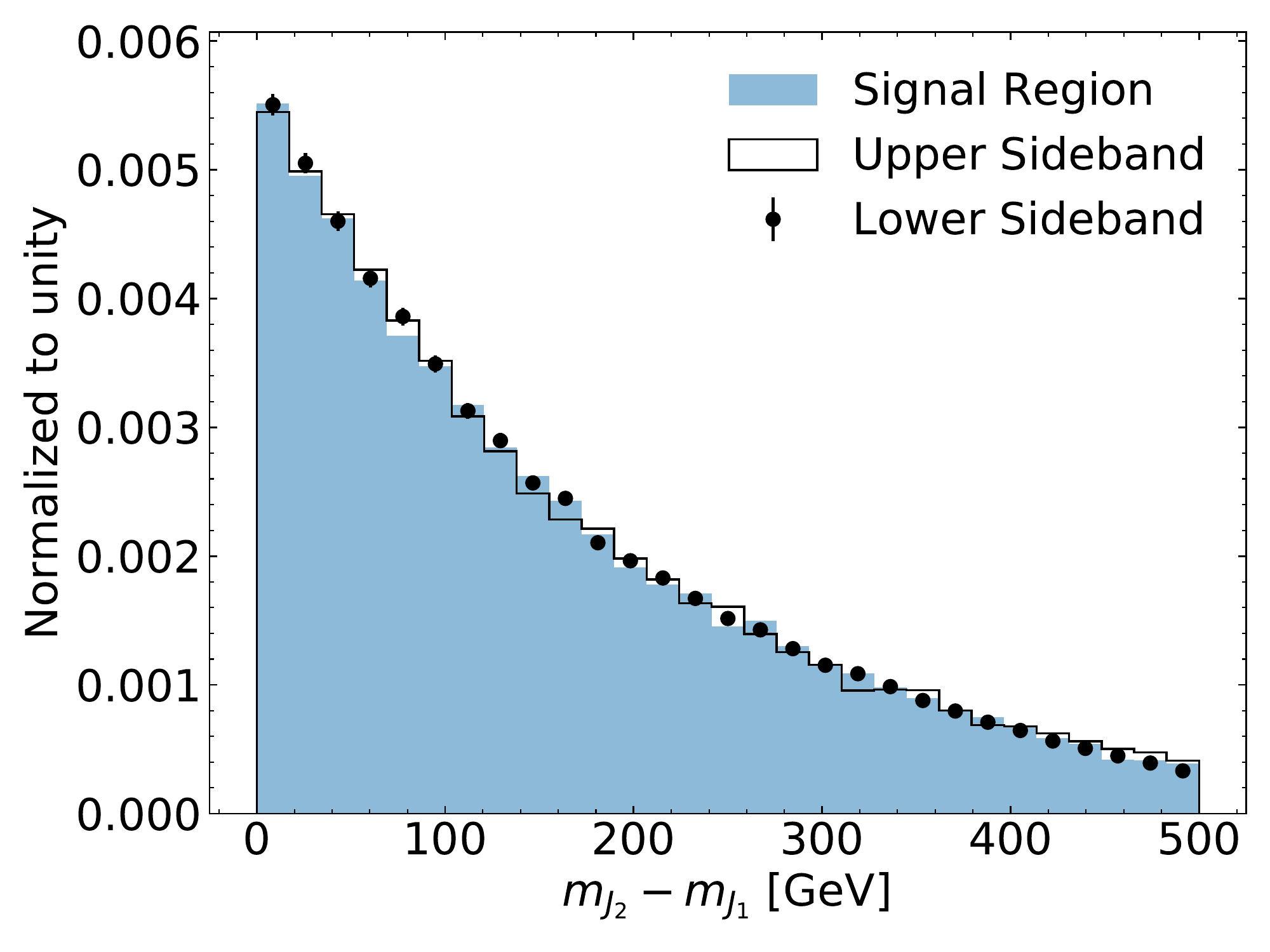}\\
\includegraphics[scale=0.35]{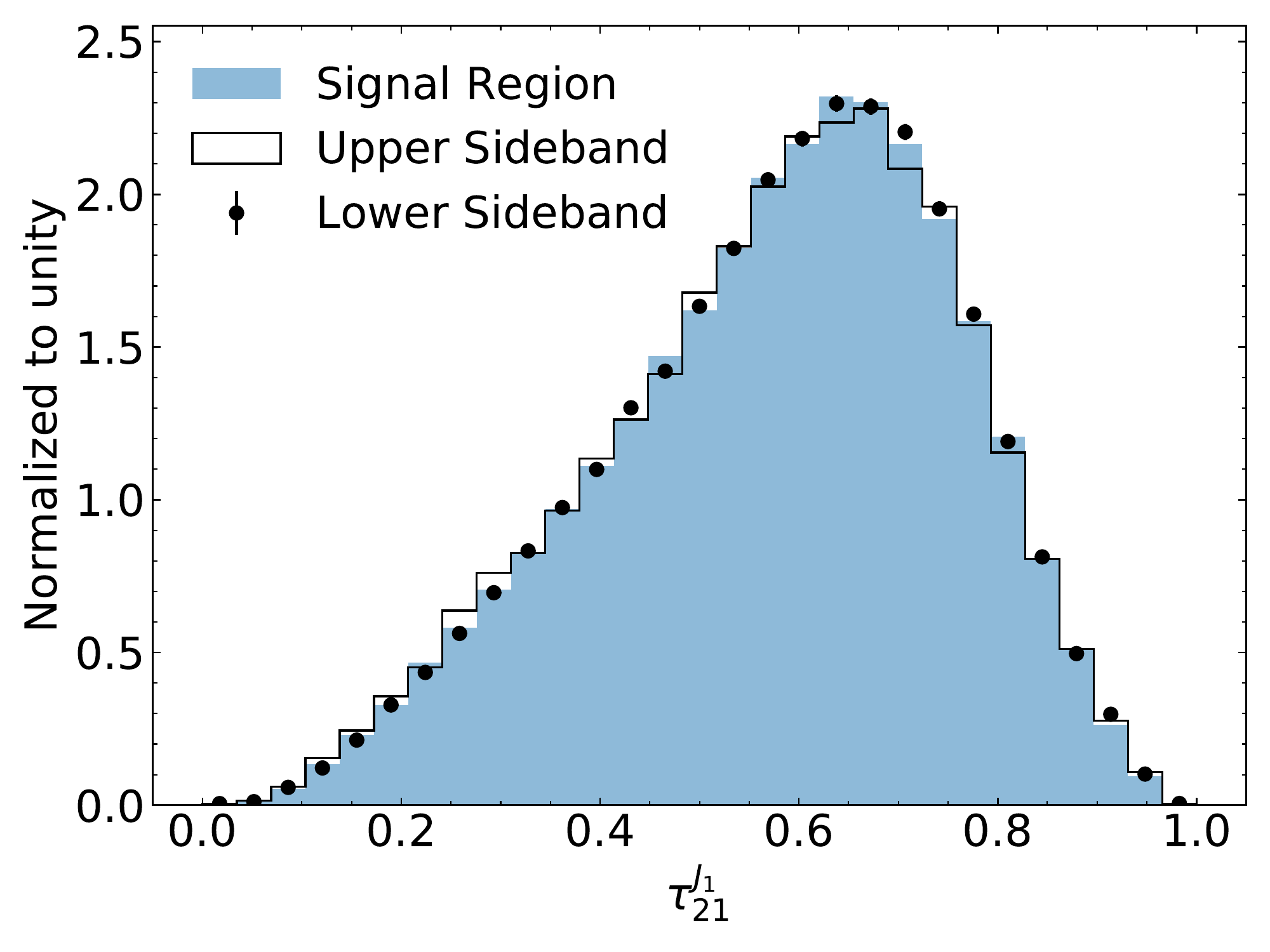}\hspace{5mm}\includegraphics[scale=0.35]{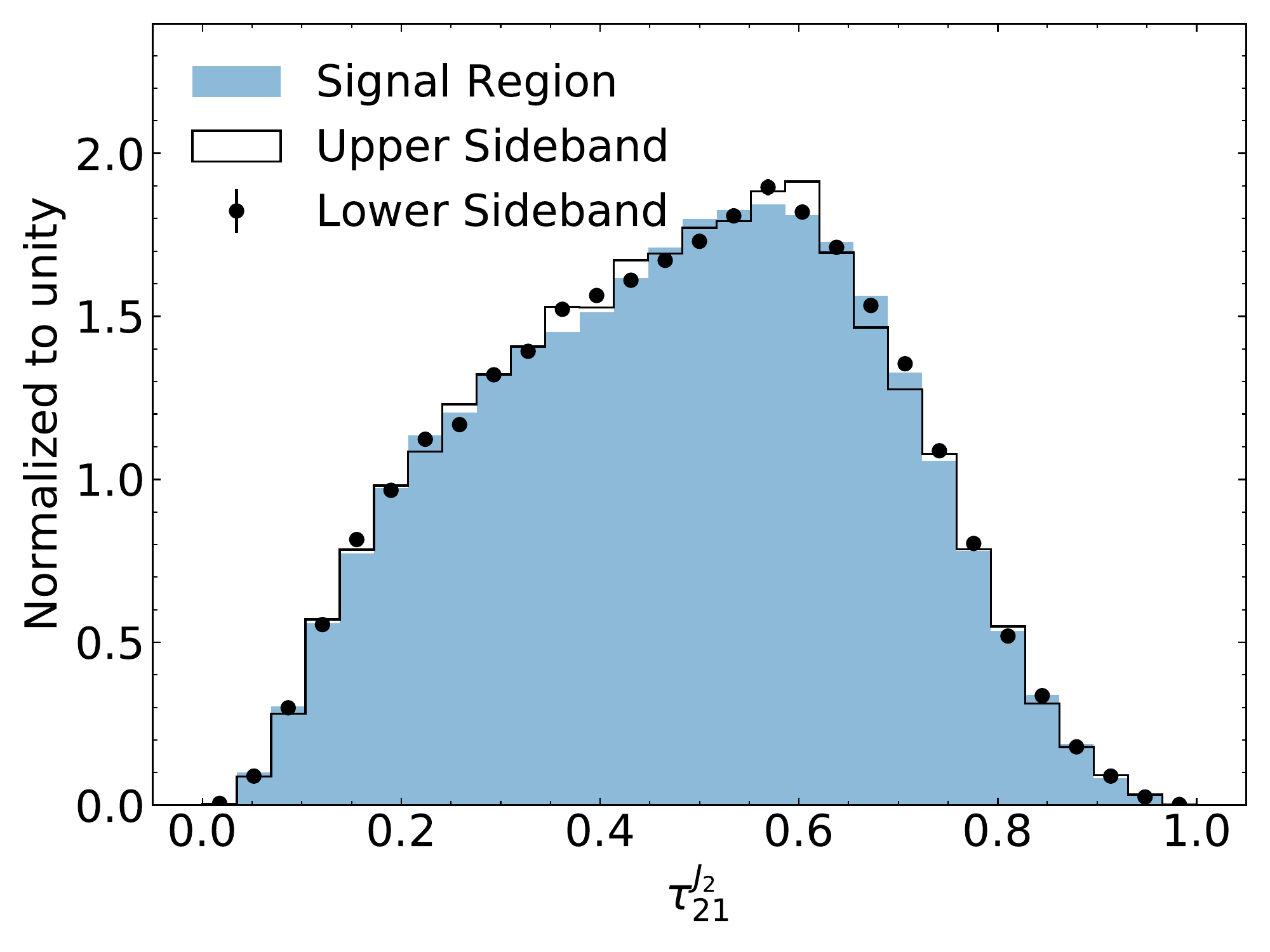}
\caption{A comparison of the four features $x$ between the SR and two nearby sidebands defined by $m_{jj}\in[3.1,3.3]$ TeV (lower sideband) and $m_{jj}\in[3.7,3.9]$ TeV (upper sideband).}
\label{fig:correlation}
\end{figure}

The results presented in the previous sections have established that ANODE is able to identify the signal and estimate the corresponding SM backgrounds introduced in Sec.~\ref{sec:sim}.  One fortuitous aspect of the chosen features $x$ introduced in Sec.~\ref{sec:sim} is that they are all relatively independent of $m_{jj}$.  This is illustrated in Fig.~\ref{fig:correlation}, using the SR and neighboring sideband regions.  As a result of this independence, the CWoLa method is able to find the signal and presumably the ANODE interpolation from SB to SR is easier than if there was a strong dependence.

\begin{figure}[h!]
\centering
\includegraphics[scale=0.55]{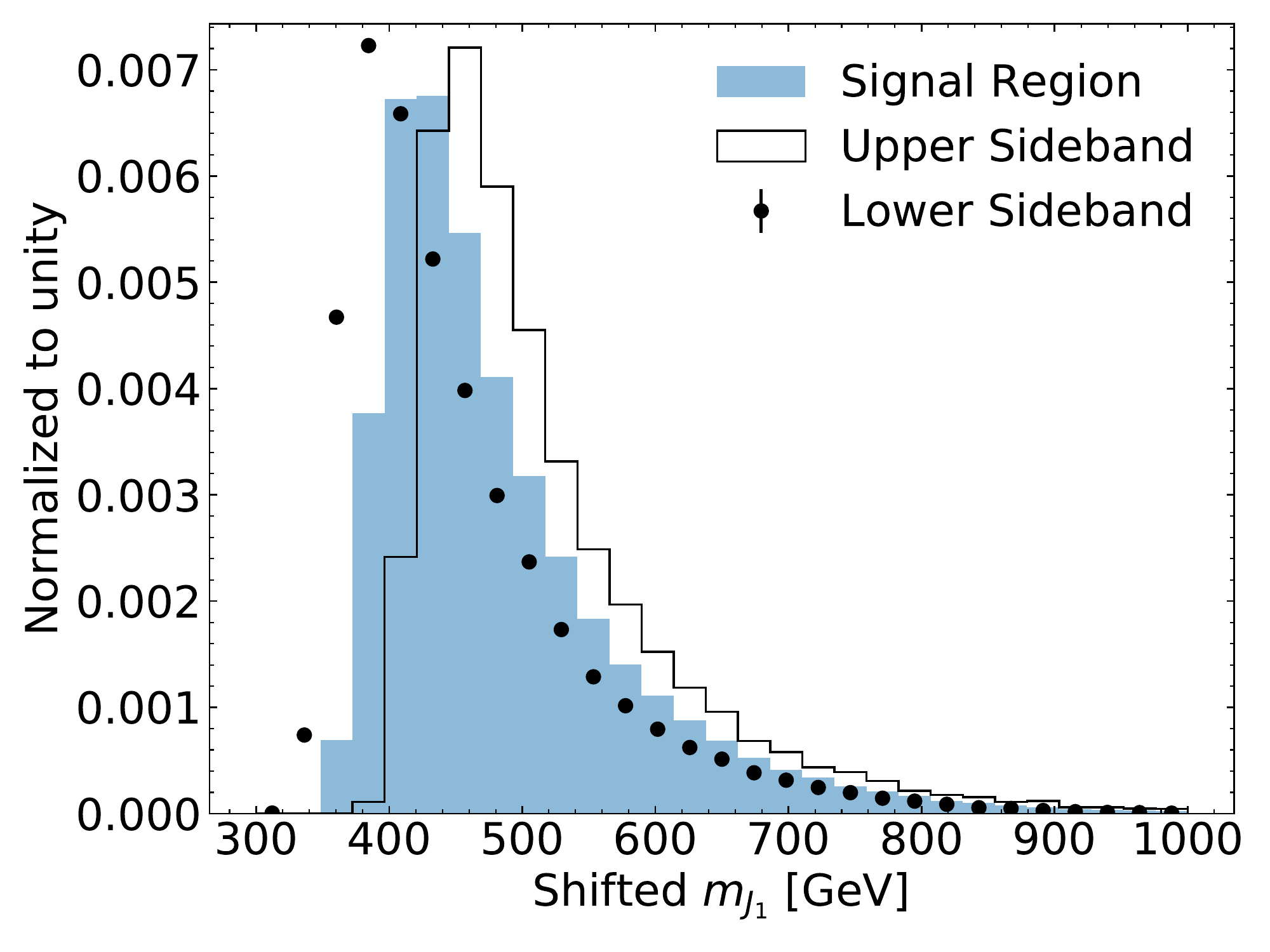}
\caption{The lighter jet mass for the SR and the lower and upper sideband regions after the shift defined by Eq.~\ref{eq:shift}.}
\label{fig:correlation2}
\end{figure}

The purpose of this section is to study the sensitivity of the ANODE and CWoLa hunting methods to correlations in the features $x$ with $m_{jj}$.  Based on the assumptions of the two methods, it is expected that with strong correlations, CWoLa hunting will fail to find the signal while ANODE should still be able to identify the presence of signal in the SR as well as estimate the background.  To study this sensitivity in a controlled fashion, correlations are introduced artificially.  In practice, adding more features to $x$ will inevitably result in some dependence with $m_{jj}$; the artificial example here illustrates the challenges already in low dimensions.   New jet mass observables are created, which are linearly shifted:
\beq
\label{eq:shift}
m_{J_{1,2}} \to m_{J_{1,2}}+c\,m_{JJ},
\eeq
where $c=0.1$ for this study.  The resulting shifted lighter jet mass is presented in Fig.~\ref{fig:correlation2}.

\begin{figure}[t!]
\centering
\includegraphics[scale=0.4]{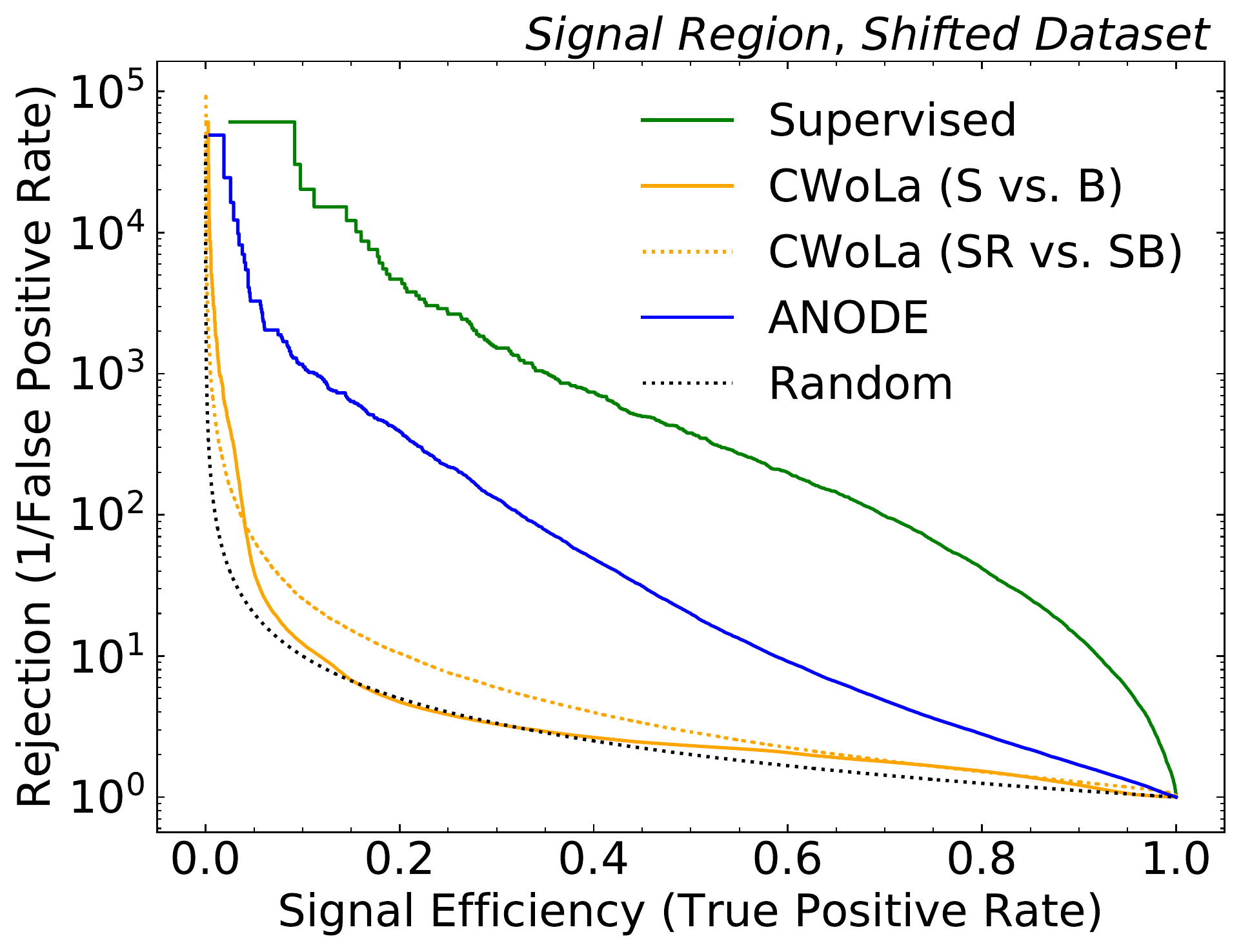}\includegraphics[scale=0.4]{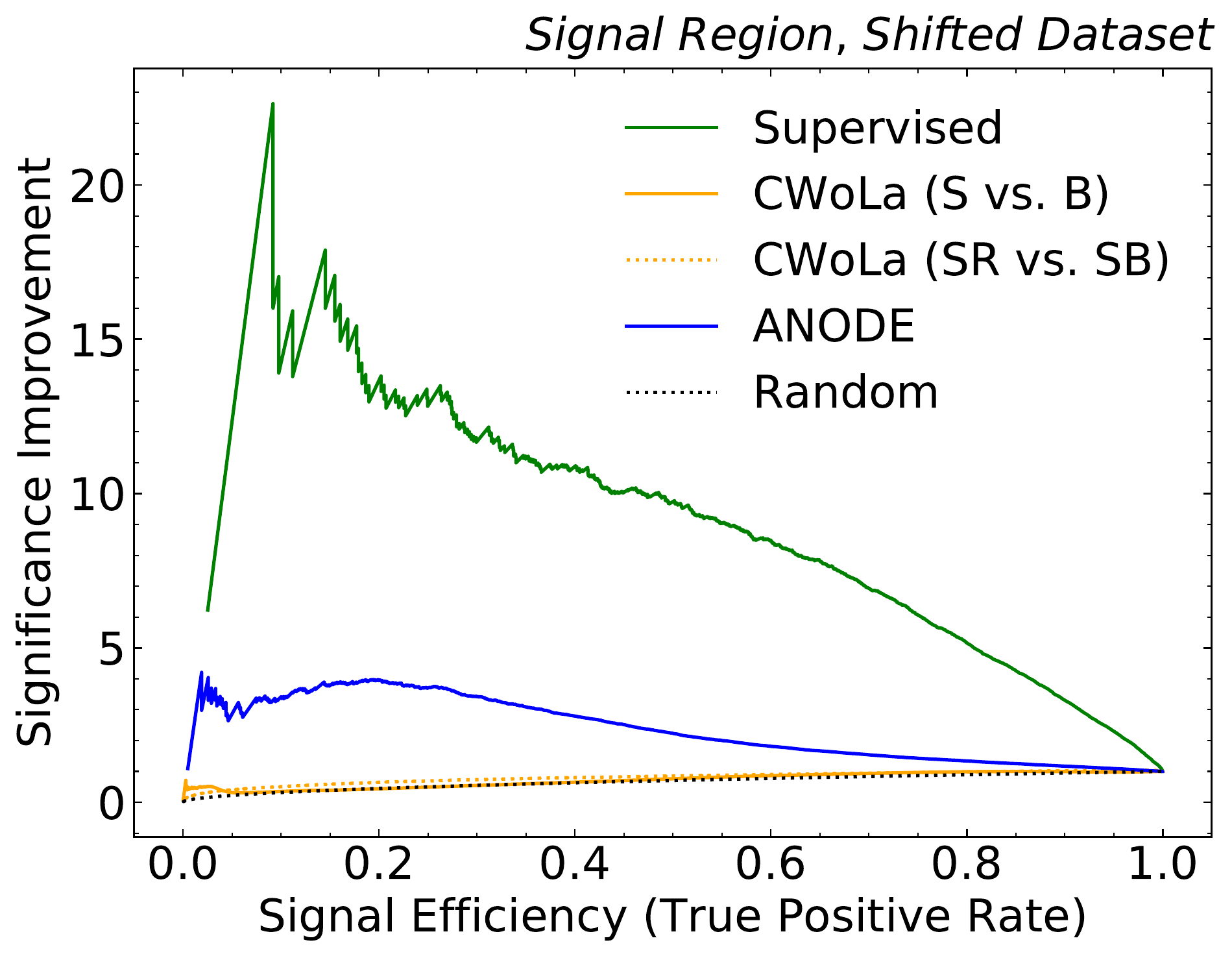}
\caption{ROC (left) and SIC (right) curves in the signal region using the shifted dataset specified by Eq.~\ref{eq:shift}.}
\label{fig:ROC_shifted}
\end{figure}

New ANODE and CWoLa models are trained using the shifted dataset and their performance is quantified in Fig.~\ref{fig:ROC_shifted}.  As expected, the fully supervised classifier is nearly the same as Fig.~\ref{fig:ROC}.  ANODE is still able to significantly enhance the signal, with a maximum significance improvement near 4.  While in principle ANODE could achieve the same classification accuracy on the shifted and nominal datasets, the performance on the shifted examples is not as strong as in Fig.~\ref{fig:ROC}.  In practice the interpolation of $p_\text{background}$ into the SR is more challenging now due to the linear correlations. This could possibly be overcome with improved training, better choices of hyperparameters, or more sophisticated density estimation techniques.

By construction, there are now bigger differences between the SR and SB than between the SR background and the SR signal.  Therefore, the CWoLa hunting classifier is not able to find the signal.  This is evident from the ROC curve in the left plot of Fig.~\ref{fig:ROC_shifted}, which shows that the signal-versus-background classifier is essentially random while the SR-versus-SB classifier has learned something non-trivial. 

\begin{figure}[h!]
\centering
\includegraphics[scale=0.35]{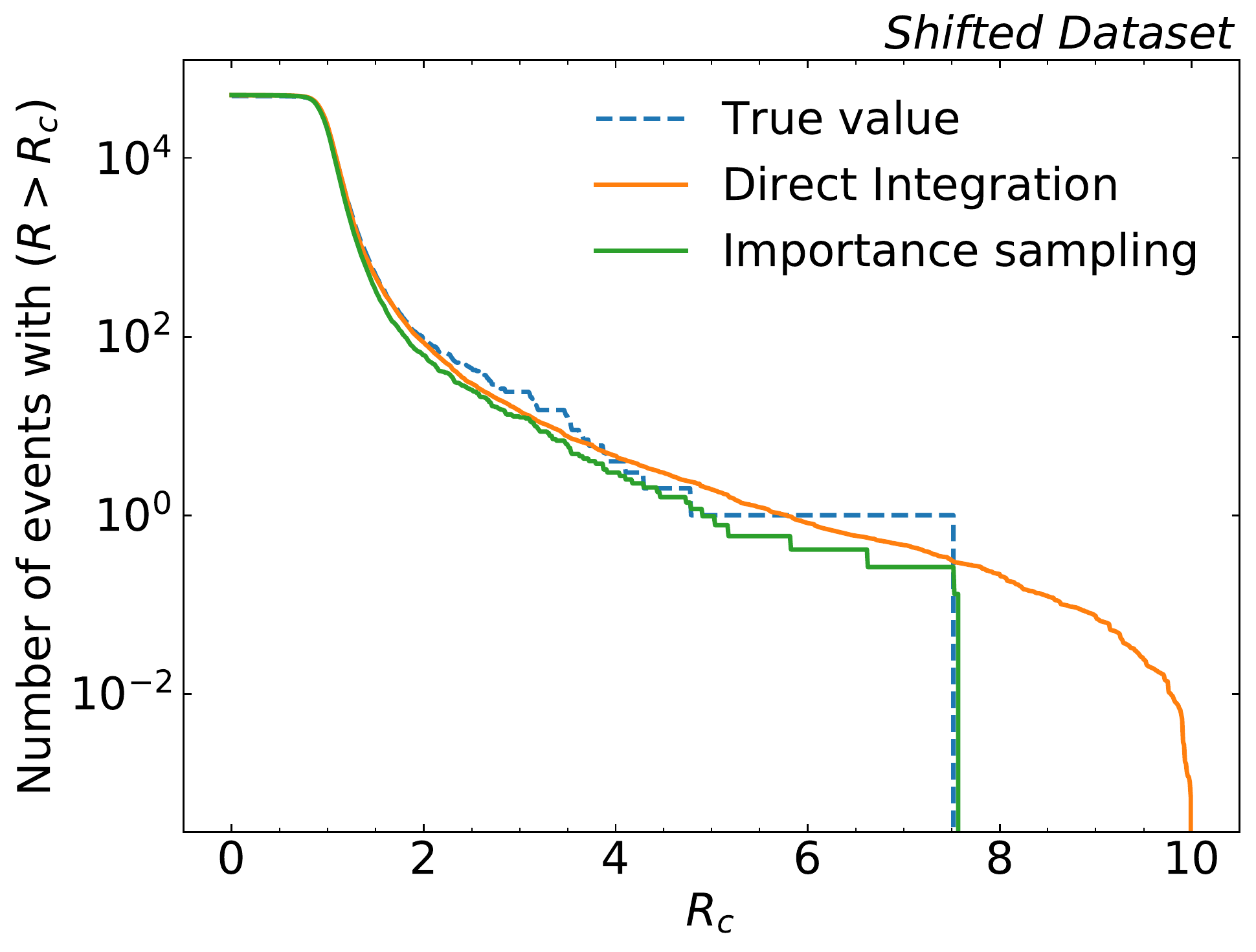}\hspace{5mm}\includegraphics[scale=0.36]{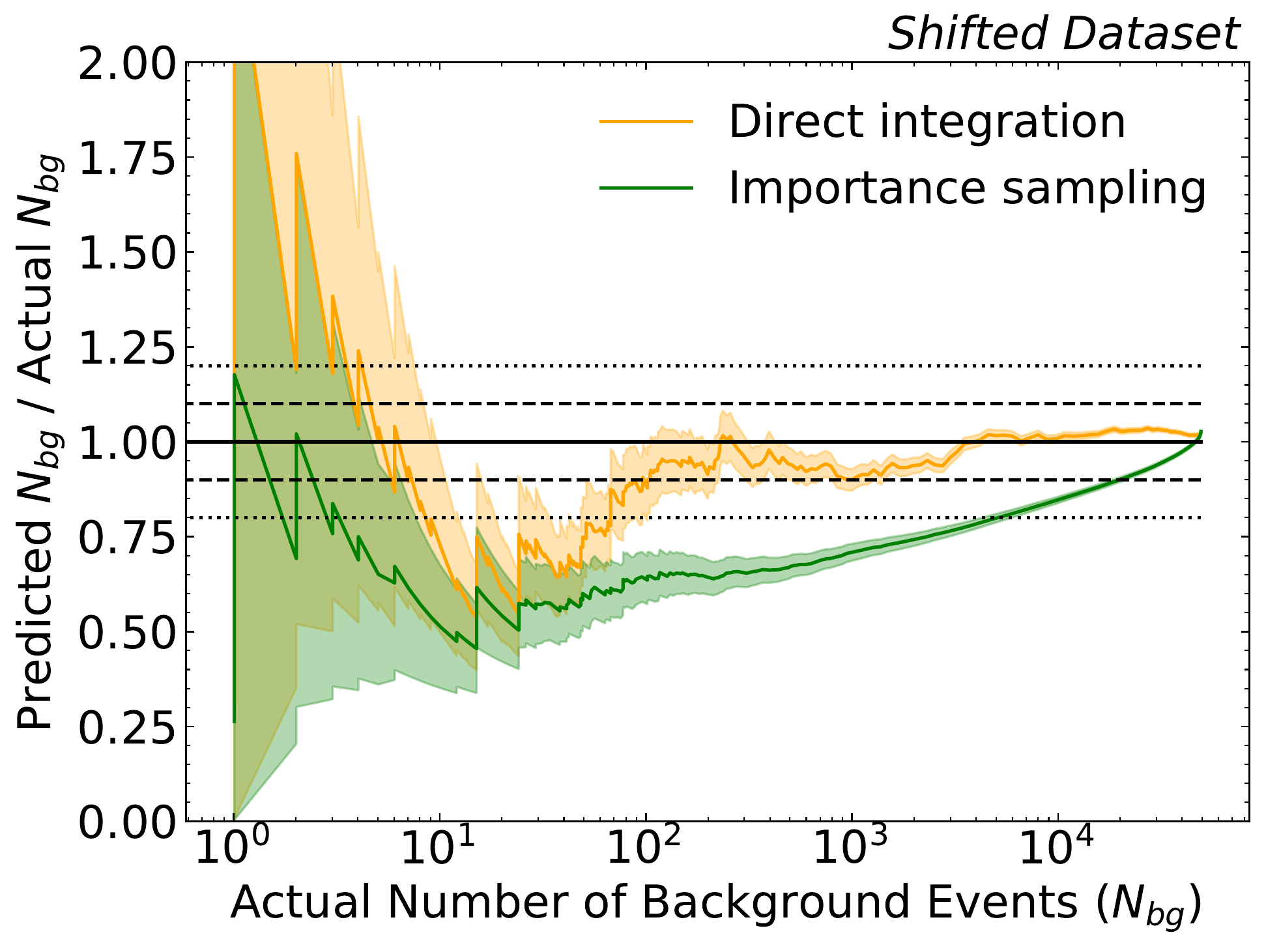}
\caption{The same as Fig.~\ref{fig:Nbgpredratio_noshift}, but for the shifted dataset.  In particular, these plots compare the background prediction from two direct density estimation techniques with the true background yield after a threshold requirement $R(x|m)>R_c$.}
\label{fig:Nbgpredratio_shift}
\end{figure}

Lastly, Fig.~\ref{fig:Nbgpredratio_shift} shows the performance of direct density estimation for the background prediction using the shifted dataset. The performance is comparable to the unshifted dataset (Fig.~\ref{fig:Nbgpredratio_noshift}), meaning that ANODE could potentially be used as a complete anomaly detection method even in the presence of correlated feature spaces.

\clearpage

\section{Conclusions}
\label{sec:conclusions}

This paper has presented a powerful new model-independent search method called \textit{ANOmaly detection with Density Estimation} (ANODE), which is built on neural density estimation.  Unlike other approaches, ANODE directly learns the background probability density and data probability density in a signal region.  The ratio of these densities is a powerful classifier and the background density can be directly used to estimate the background efficiency from a threshold requirement on the classifier.  Finally, ANODE is robust against correlations in the data, which tend to break other model-agnostic sideband methods such as CWoLa. 

The results presented in this paper are meant to be a proof of concept of the general method, and there are many exciting future directions. For example, while this paper focused on collider searches for BSM,  the ANODE method is completely general and could be applied to many areas beyond high energy physics, including astronomy and astrophysics. Similarly, while the demonstrations here were based on the innovative MAF density estimation technique, the ANODE method can be used in conjunction with any density estimation algorithm.  Indeed, there are numerous other neural density estimation methods from the past few years that claim state-of-the-art performance, including Neural Autoregressive Flows \cite{DBLP:journals/corr/abs-1804-00779} and Neural Spline Flows \cite{durkan2019neural}; exploring these would be an obvious way to attempt to improve the results in this paper. In addition, it would be interesting to attempt the ANODE method on even higher-dimensional feature spaces, all the way up to the full low-level feature set of the four vectors of all the hadrons in the event.   This might already be feasible with existing neural density estimators, at is it common to evaluate their performance on high dimensional datasets ranging from UCI datasets~\cite{Dua:2019} with up to $\sim50$ features, to image datasets such as MNIST~\cite{lecun2010mnist} and CIFAR-10~\cite{cifar-10} which have hundreds and thousands of features respectively.  The prospects for the ANODE method are exciting: as the field of neural density estimation continues to grow within the machine learning community, ANODE  will become more sensitive to resonant new physics in collider high energy physics and beyond.

\acknowledgments

DS is grateful to Matt Buckley and John Tamanas for many fruitful discussions on neural density estimation. We are especially grateful to John Tamanas for help with the conditional MAF code.   Additionally, we would like to thank Kyle Cranmer and Uro\v{s} Seljak for helpful discussions and Nick Rodd and John Tamanas for helpful comments on the draft.  This work was supported by the U.S.~Department of Energy, Office of Science under contract DE-AC02-05CH11231. DS is supported by DOE grant DOE-SC0010008.  DS thanks LBNL, BCTP and BCCP for their generous support and hospitality during his sabbatical year. 

\appendix

\section{Comments on optimality}
\label{sec:optimal}

The Neyman-Pearson lemma only applies to simple hypothesis tests.  The lemma states that for a fixed probability of rejecting the null hypothesis when it is true (level), the probability for rejecting the null hypothesis when the alternative is true (power) is maximized with the likelihood ratio test statistic.  For supervised searches with profiled nuisance parameters or for anomaly detection with a composite alternative hypothesis, there is no uniformly most powerful classifier.  The goal of this brief section is to clarify what is meant by \textit{asymptotically optimal anomaly detection}.

For any given BSM model, the procedures labeled asymptotically optimal are likely not optimal.  The sense in which they are optimal is as follows.  Let the null hypothesis $H_0$ be that the data are distributed according to $p_\text{background}$, a density describing the phase space of the background-only.  Furthermore, let the alternative hypothesis $H_A$ be that the data are distributed according to $p_\text{data}$, the learned density of the data.  Distinguishing $H_0$ from $H_A$ is a simple hypothesis test.  Therefore, the test statistic $p_\text{background} / p_\text{data}$ has the property that for a fixed probability for rejecting $H_0$ given $\text{data}\sim p_\text{background}$, the probability for rejecting $H_0$ is as high as possible when $H_A$ is true (which it is).  If $p_\text{background} = p_\text{data}$, then power = level.   So ANODE is asymptotically optimal for rejecting the data as background-only, but is not `optimal' for rejecting any particular BSM model.

\bibliographystyle{jhep}
\bibliography{myrefs}
\end{document}